%% file: main.tex
\pdfoutput=1

\documentclass[11pt]{article}

\usepackage[preprint]{acl}

\usepackage{times}
\usepackage{latexsym}

\usepackage[T1]{fontenc}

\usepackage[utf8]{inputenc}
\usepackage{multirow}
\usepackage{booktabs}
\usepackage{microtype}

\usepackage{inconsolata}

\usepackage{graphicx}

\usepackage{tabularx}
\usepackage{makecell}
\usepackage{colortbl}         

\usepackage{amsmath}

\usepackage{hyperref}       
\usepackage{url}            
\usepackage{amsfonts}       
\usepackage{nicefrac}       
\usepackage{microtype}      
\usepackage{amsthm}
\usepackage{tabularx}
\usepackage{makecell}
\usepackage{algorithm}
\usepackage{algorithmic}
\usepackage{multirow}
\usepackage{pdfpages}
\usepackage{subfigure}
\usepackage{bbding}
\usepackage{multibib}
\usepackage{array}
\usepackage{listings}
\usepackage{setspace}

\usepackage{wrapfig}

\definecolor{cvprblue}{rgb}{0.21,0.49,0.74}

\date{}

\newtheorem{proposition}{Proposition}

\newcommand{\INPUT}{\STATE \textbf{Input:}}

\title{``Yes, My LoRD.'' Guiding Language Model Extraction with\\
  Locality Reinforced Distillation}

\author{{ Zi Liang$^{\dagger}$} { Qingqing Ye$^{\dagger}$} {
    Yanyun Wang$^{\flat}$} { Sen Zhang$^{\dagger}$} 
   { Yaxin Xiao$^{\dagger}$}\\ \textbf{ Ronghua Li$^{\dagger}$} \textbf{
     Jianliang Xu$^{\ddagger}$} \textbf{ Haibo
     Hu$^{\dagger}$\thanks{Corresponding author.}}
  \\
$\dagger$: The Hong Kong Polytechnic University, Hong Kong, China\\
$\flat$: The Hong Kong University of Science and Technology
(Guangzhou), China\\
$\ddagger$: Hong Kong Baptist University, Hong Kong, China\\
\texttt{\{zi1415926.liang,20034165r,22041986r\}@connect.polyu.hk}\\
\texttt{\{qqing.ye,senzhang,haibo.hu\}@polyu.edu.hk}\\
\texttt{ ywang856@connect.hkust-gz.edu.cn, xujl@comp.hkbu.edu.hk} 
}

\begin{document}

\maketitle

\begin{abstract}
Model extraction attacks (MEAs) on large language models (LLMs) have received increasing attention in recent research. However, existing attack methods typically adapt the extraction strategies originally developed for deep neural networks (DNNs). They neglect the underlying inconsistency between the training tasks of MEA and LLM alignment, leading to suboptimal attack performance. To tackle this issue, we propose \underline{Lo}cality \underline{R}einforced
\underline{D}istillation (LoRD), a novel model extraction algorithm
specifically designed for LLMs. In particular, LoRD employs a newly
defined policy-gradient-style training task that utilizes the
responses of victim model as the signal to guide the crafting of
preference for the local model. Theoretical analyses demonstrate that
\uppercase\expandafter{\romannumeral1}) The convergence procedure of
LoRD in model extraction is consistent with the alignment procedure of
LLMs, and \uppercase\expandafter{\romannumeral2}) LoRD can reduce
query complexity while mitigating watermark protection through our
exploration-based stealing. Extensive experiments validate the
superiority of our method in extracting
various state-of-the-art commercial LLMs. Our code is available at: \url{https://github.com/liangzid/LoRD-MEA}.
\end{abstract}

\input{intro}

\input{background}
\input{method}
\input{analysis}
\input{exper}

\section{Conclusion}

In this paper, we have focused on the extraction problem of commercial
large language models. We proposed LoRD, a practical and realistic
extraction algorithm which is consistent with the alignment procedure
of large language models. Our analysis proved that LoRD can reduce the
query time significantly and mitigate the certification of current
watermarks naturally, surpassing existing MEA algorithms' capabilities. Extensive experiments on domain-specific
stealing and alignments demonstrated the superiority of our method.

\section*{Acknowledgment}
The authors would like to thank the reviewers for their detailed suggestions.
This work was supported by the National Natural Science Foundation of
China (Grant No: 92270123 and 62372122), the Research Grants
Council, Hong Kong SAR, China (Grant No: 15203120, 15209922, 15210023,
15224124, and C2004-21GF), and the Innovation and Technology Fund
(Grant No: ITS-140-23FP).

\section*{Limitations and Future Works}
\noindent
\textbf{MEAs on Multi-modal Models.} While this paper delves into MEAs
for large language models, it acknowledges the oversight of the multi-modal attribution of current commercial models~\citep{gemini,gpt4} that integrate various forms of data such as text, images,
voice, and so on. The challenge of extending MEA algorithms to
accommodate these models, which requires extra considerations on the
unified representation of concepts, remains unexplored. Future work
could focus on developing MEA methodologies sensitive to multi-modal data nuances.

\noindent
\textbf{Capacities beyond LaViSH Settings.} We utilize the LaViSH
setting to describe the model capacity of adversaries in our threat
model (see Appendix \ref{sec:threat}). However, sometimes, the adversary
might possess comparable or superior training resources to the
victims. Though this paper posits that our MEA algorithms and
theoretical analysis are still compatible with such conditions, we
concede that concrete experimental validation and results beyond
LaViSH settings are not presented here.

\noindent
\textbf{Lower-level Extractions.}
This study evaluates MEAs at the performance level, i.e., it measures the
extraction effectiveness simply through task performance
metrics, or the similarity of learned distributions to the victim model.
This setting is justified, as performance metrics are
essential for evaluating task-related knowledge and the practical
application of LLMs. However, it does not consider the lower-level
similarities between the victim and local models.
Can we achieve neuron-level alignments in LLM's MEAs? How does a LaViSH
setting hurt LLM's MEAs? Is it compatible to extract a MoE
(Mix-of-the-Expert)~\citep{moe} victim model with a dense local model? These
questions are not addressed in this research.

\section*{Ethical Considerations}
As discussed in Section \ref{sec:intro}, MEAs
are becoming increasingly prevalent in industrial settings and have
already been executed, yet there remains a critical gap in
understanding which specific tasks are more susceptible and what
capabilities are necessary for effective executions. This lack of
knowledge exacerbates the challenges faced by LLM maintainers in
safeguarding their systems. Our research can contribute
to that. Besides, the theoretical problem we
address (as shown in Section \ref{sec:th-an}) offers a novel and
insightful perspective on the nature of this threat. Based on
  these two points, \textbf{we believe the benefits of our paper
    outweigh potential harms, which aligns with the
  principles of the \emph{Menlo Report}~\citep{menlo} on ethics.}

 Additionally, we have submitted an
  anonymous version of the paper to the maintainers of the
  victim models used in our study to assist in improving their model security.

It is important to acknowledge, however, that the algorithms we
propose could inadvertently enhance the efficiency of illicit
extraction efforts by adversaries. To mitigate this risk, we have
introduced and analyzed two defensive strategies, assessing
both their effectiveness and potential vulnerabilities under adaptive
attack scenarios.

Potential defenses consist of:


$\bullet$ \textbf{Query Detection.} One approach to effectively prevent the
attack of LoRD is by detecting the distribution of query
texts. This is because LoRD, similar to current MEA algorithms, makes
no improvements to query samples, indicating that it can be
detected by analyzing the statistical information of the adversary's
queries, such as the number of queries, distribution of query
contents, and so on.
However, this defense is usually resource-consuming, as it requires
the LLM provider to store all query texts of each user. Besides, the
potential for false positives could adversely affect the
user experience.

$\bullet$ \textbf{More Powerful Watermarks.} While we highlight the watermark
resistance of LoRD, watermarking remains one of the most effective
solutions to mitigate MEAs. For example, some model-level watermarks,
such as backdoor-based watermarking~\citep{backdoor-mea-wm,mea-defender}, can effectively certify
the theft of DNNs. While model-level (e.g. backdoor-based) watermarks
on pre-trained models raised increasing concerns recently~\citep{model-wm1,model-wm2,model-wm3}, model-level watermarking on LLMs remains preliminary. Besides, this technique might not work when the
adversary only steals a subset of knowledge in
which no backdoor is embedded.

\clearpage
\bibliography{refs}

\newpage

\appendix

\input{appendix}

\end{document}

%% file: intro.tex
\section{Introduction}\label{sec:intro}

In recent years, we have witnessed the remarkable success of large language models (LLMs) such as ChatGPT~\citep{gpt4}, Gemini~\citep{gemini}, and Claude~\citep{claude}, which are now widely employed in various consumer and industrial applications.
Despite their success, these models may suffer from \emph{model extraction attacks} (MEAs)~\citep{mea-bertapi,mea-vendor,mea-code}, where their knowledge could be at risk of being stolen by an adversary through a \emph{local model} that learns on the data collected from the \emph{victim model}.
Besides of some ``open-source'' LLMs (e.g., Alpaca~\citep{alpaca}),
which are trained on the chat history of GPT-4, cases of commercial model theft among companies have also been reported recently~\citep{company-steal}.

\begin{figure}
  \centering
\includegraphics[width=0.85\linewidth]{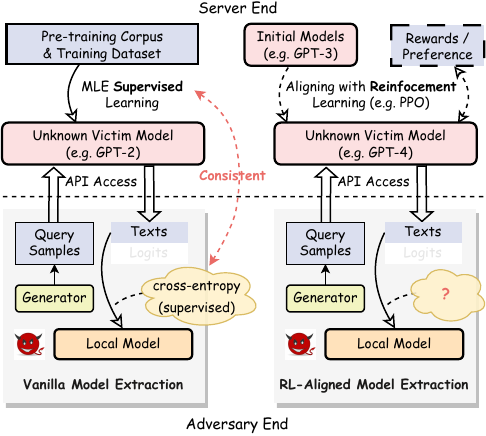}
\caption{
Comparison between vanilla MEAs on conventional DNNs (left) and MEAs
on LLMs with alignments (right).
}
\label{fig:intro}
\end{figure}

Under such a real-world threat, instead of focusing on MEAs against conventional DNNs, which have been extensively studied theoretically~\citep{mea-the-dyn,mea-the-stu,mea-the-a} and empirically~\citep{mea-tra-high,mea-tra-steal,mea-tra-prac,mea-11,mea-12,mea-13}, a few recent works turn to explore model extraction algorithms and theorems for LLMs.
For example, \citet{mea-wmt} propose a monolingual-query-based imitation attack framework to steal machine translation knowledge from generative language models such as GPT-2. \citet{mea-code} investigate threats of stealing the code-related knowledge from LLMs.
However, these studies inherit those MEA
algorithms from traditional fields, such as computer vision~\citep{mea-tra-steal,mea-tra-prac}, and train the local
model via supervised learning like maximum likelihood estimation (MLE)~\citep{bengio2000neural,mle}, while neglecting the inconsistency of training tasks between MEAs and the alignments~\citep{instructGPT,sparrow,rlhf1,rlhf2,rlhf3} of modern LLMs. As shown in Figure \ref{fig:intro}, modern LLMs typically employ alignments using reinforced learning, which is missing in the local model training of conventional MEAs. As a result, these attacks usually suffer from poor performance.

In this paper, we challenge the effectiveness of MLE in stealing a
reinforcement-learning-aligned LLM, by analyzing its following
drawbacks: \emph{i) Low query efficiency.} Current
MEAs on LLMs suffer from unacceptably
significant query times because they must collect enough generated
responses, which entails exponential complexity in terms of generated
tokens, resulting in low query efficiency.
\emph{ii) Vulnerability against defenses.} Directly learning from the
responses of victim models can cause local models
to inadvertently incorporate those
\emph{watermarks}~\citep{wm1,wm2,wm3,wm4} embedded in the output of victim
models. The residue of such watermarks makes the extraction
less stealthy and even serves as provenance evidence of model theft.

Motivated by these limitations, we propose \underline{Lo}cality
\underline{R}einforced \underline{D}istillation (LoRD), a
query-efficient and watermark-resistant model extraction attack under
a training paradigm similar to LLM's alignments.
Stealing LLMs via reinforcement learning (RL) paradigms is challenging. The
main reason is that the alignment procedure of LLMs heavily relies on the feedback
signal of \emph{\textbf{human annotators}}~\citep{rlhf1,rlhf2,rlhf3}, which is difficult to reproduce directly in the context of MEAs.
To tackle this challenge, we develop a
policy-gradient-style extraction procedure. This approach regards the
\emph{locality direction} between the generations of local models and
victim models as the implicit reward signal. It can thus achieve a \emph{\textbf{human-feedback-free}} RL for our attack.
From the theoretical perspective, we show why those existing MEAs using \emph{MLE} and \emph{knowledge distillation (KD)} are inconsistent with the optimization procedure in LLMs' alignments. Along this way, we also demonstrate why LoRD can achieve stronger watermark resistance and higher query efficiency.

Extensive experiments on five downstream NLP tasks and two alignment tasks with 12 datasets
demonstrate that it is feasible to steal a commercial LLM with 175
billion parameters by a pre-trained local model with only 8 billion
parameters under a given domain. The resulting local model performs
statistically similar to the
victim model for tasks not requiring extra knowledge (e.g., data-to-text),
and only $0\sim 3$ percentage lower for tasks requiring it (e.g.,
translation and QAs). This result poses an immediate threat of
task-specific and alignment extraction on commercial LLMs.
Our contribution are summarized as follows:


\noindent $\bullet$~\textbf{New Perspective of Large Language Model Extraction.} We present LoRD, a novel model extraction attack algorithm for LLMs.
To our best knowledge, it is the first effective and realistic
extraction algorithm that takes LLM alignment into consideration for MEAs.

\noindent $\bullet$~\textbf{Theoretical Guarantee.} We theoretically prove that the
  convergence procedure of LoRD in MEAs is consistent with the
  alignments of LLMs. Furthermore, we demonstrate that LoRD can reduce
  query complexity while mitigating watermark protection through exploration-based stealing.

\noindent $\bullet$~\textbf{Systematical Evaluation.} Extensive experiments demonstrate that our method outperforms current extraction strategies across different downstream NLP tasks.

%% file: background.tex
\section{Background}\label{sec:bg}


\subsection{Policy Gradient Models}\label{sec:pgm}
Policy gradient models (PGM) are commonly used in reinforcement
learning (RL) algorithms to optimize the agents based on the decided
\emph{action} of RL agents. Represented by TRPO~\citep{trpo} and
PPO~\citep{ppo}, policy gradient models minimize the
the following objective function:
\begin{equation}
\label{eq:pg}
\mathcal{L}_{pg,j}=- \hat{\mathbb{E}}_{j}[p_{j}^{r}(\theta)A_{j}],
\end{equation}
where at each decision step $j$, $p_{j}^{r}(\theta)=\frac{\pi_{\theta}(a_{j}|s_{j})}{\pi_{\theta_{old}}(a_{j}|s_{j})}$ refers to the probability ratio defined by
the optimized policy $\pi_{\theta}(a_{j}|s_{j})$ and the initial
policy $\pi_{\theta_{old}}(a_{j}|s_{j})$,
$s_{j}$ denotes the \emph{state} of the environment, $a_{j}$ denotes
the decided \emph{action} of $\pi_{\theta}$, and
$A_{j}$ is the \emph{de-biased reward} of $a_{j}$. $A_{j}$ is
estimated by the $Q$-value minus the $V$-value, i.e.,
\begin{equation}
\label{eq:rl-reward}
A_{j}(s_{j},a_{j})=Q(s_{j},a_{j})-V(s_{j}).
\end{equation}

Intuitively, $Q$-value refers to the \emph{reward} if employing action
$a_{j}$ at the given environment state $s_{j}$, which can be seen as
the label of policy's decision. $V$-value represents the estimation of
the expected reward at $s_{j}$. Consequently, $A_{j}$ denotes the
\emph{surprise} when taking action $a_{j}$.


\begin{figure*}[t]
\includegraphics[width=0.99\linewidth]{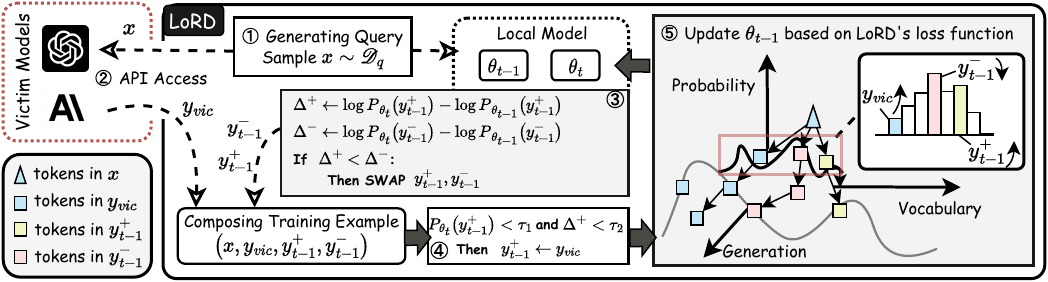}
\caption{The stealing procedure of LoRD.
  }
\label{fig:lord}
\end{figure*}

\subsection{Language Modeling}\label{sec:bg-lm}
\noindent
\textbf{Supervised Training (SFT).}
Given a pre-trained model with parameters $\theta$, supervised
training is essentially the \emph{maximum likelihood estimation (MLE)}
task~\citep{bengio2000neural,mle}, which fine-tunes $\theta$ on the
labeled dataset
$\mathcal{D}_{tr}^{s}=\{(\mathbf{x}_{i},\mathbf{y}_{i})|i=1,2,...,N_{trs}\}$
by minimizing the following objective function:
\begin{equation}\small
\label{eq:mle}
\mathcal{L}_{mle}= -\prod_{i}^{N_{trs}}{P_{\theta}(\mathbf{y}_{i}|\mathbf{x}_{i})}=-\prod_{i}^{N_{trs}}\prod_{j}^{N}{P_{\theta}({y}_{i,j}|\mathbf{x}_{i},\mathbf{y}_{i,<j})},
\end{equation}
where $N$ denotes the sequence length of $\mathbf{y}_{i}$, $y_{i,j}$
denotes the $j$-th token in $\mathbf{y}_{i}$, and $\mathbf{y}_{i,<j}=\{y_{i,0},...,y_{i,j-1}\}$. The logarithmic formula of
Equation \ref{eq:mle}
can also be seen as a \emph{joint cross-entropy} loss function:
\begin{equation}\small
\label{eq:ce}
\begin{aligned}
\mathcal{L}_{ce}&=- \sum_{i}^{N_{trs}}{\text{log}P_{\theta}(\mathbf{y}_{i}|\mathbf{x}_{i})}\\&=-\sum_{i}^{N_{trs}}\sum_{j}^{N}{\text{log}P_{\theta}({y}_{i,j}|\mathbf{x}_{i},\mathbf{y}_{i,<j})}.
\end{aligned}
\end{equation}



\noindent
\textbf{Aligning from Preferences.}
Employing reinforcement learning in LLMs typically
consists of three stages. First, the annotators construct a preference dataset
$\mathcal{D}^{pref}=\{(\mathbf{x}_{i},\mathbf{y}_{i}^{+},\mathbf{y}_{i}^{-})\}$
by chatting with LLMs and rating their responses,
where $\mathbf{y}_{i}^{+}$ and $\mathbf{y}_{i}^{-}$ denote the rated
positive and negative responses of the dialogue context
$\mathbf{x}_{i}$, respectively. Then, a \emph{reward model}
$R_{\theta_{\phi}}(\mathbf{x},\mathbf{y})\rightarrow \mathbf{r}$ is
trained based on $\mathcal{D}^{pref}$ to simulate the environment and
predict the reward values of tokens in given texts. It is trained with a
pair-wise loss,
\begin{equation}\small
\label{eq:r}
\mathcal{L}_{r}=-\sum_{(\mathbf{x},\mathbf{y}^{+},\mathbf{y}^{-}) \sim
\mathcal{D}^{pref}}{\sigma(R_{\theta_{\phi}}(\mathbf{x},\mathbf{y}^{+})-R_{\theta_{\phi}}(\mathbf{x},\mathbf{y}^{-}))},
\end{equation}
where $\sigma(\cdot)$ denotes the sigmoid function.
Based on the reward model $R_{\theta_{\phi}}(\mathbf{x},\mathbf{y})$, we can
finally train the language models $P_{\theta}$ by maximizing its reward:
\begin{equation}\small
\label{eq:align}
\max\limits_{\theta}{\sum_{\mathbf{x} \sim \mathcal{D}_{q}}{R_{\theta_{\phi}}(\mathbf{x},\hat{\mathbf{y}})}}-\beta\mathbb{D}_{KL}[P_{\theta}(\hat{\mathbf{y}}|\mathbf{x})||P_{\theta_{init}}(\hat{\mathbf{y}}|\mathbf{x})],
\end{equation}
where $\mathcal{D}_{q}$ denotes the dataset of text inputs,
$\hat{\mathbf{y}} \sim P_{\theta}(\mathbf{y}|\mathbf{x}))$ denotes the
sampled sequence of the training model, and
$\theta_{init}$ is the initialized parameters of the model, e.g., the
parameters after SFT. The Kullback-Leibler (KL) divergence term,
$\beta\mathbb{D}_{KL}[P_{\theta}(\mathbf{y}|\mathbf{x})||P_{\theta_{init}}(\mathbf{y}|\mathbf{x})]$,
introduced by TRPO~\citep{trpo},
is incorporated to constrain the shift of distribution in generated texts
$\hat{\mathbf{y}}$, where $\beta$ is the hyperparameter.

Consequently, SFT shown in Equation \ref{eq:ce} fine-tunes the
pre-trained model with parameters $\theta_{pre}$ into an aligned model $\theta_{sft}$ through MLE, and RLHF outlined in
Equation \ref{eq:align}, further aligns $\theta_{sft}$ towards the
target model $\theta_{vic}$. As this procedure is not consistent
  with the conventional training framework of DNNs, it remains unclear whether current MEAs (detailed in Appendix
\ref{sec:related-MEA}) are effective and efficient in stealing a
LLM. Specifically, we will first put forward a new stealing method in
Section \ref{sec:lord}, and compare it with current MEAs in Section \ref{sec:th-an}.


%% file: method.tex
\section{LoRD: Locality Reinforced Distillation}\label{sec:lord}


\subsection{Overview}\label{sec:overview}
\begin{figure}
\includegraphics[width=\linewidth]{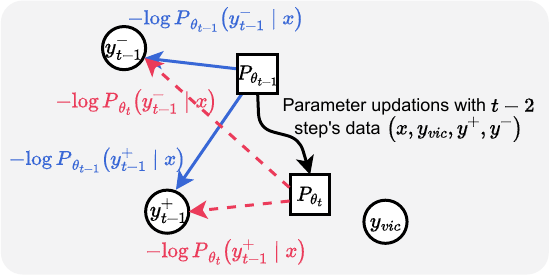}
\caption{\small {Determination of the positive and
    negative samples in LoRD.} We sample
  $\mathbf{y}_{t-1}^{+}$ and $\mathbf{y}_{t-1}^{-}$ from
  $P_{\theta_{t-1}}(\cdot |\mathbf{x})$, and compute their conditional
  probabilities. The response with a higher probability increment on $\theta_{t}$ is selected as the
  positive sample.
}
\label{fig:explain}
\end{figure}
In this subsection, we delve into the details of our model extraction
framework, LoRD (\underline{Lo}cality \underline{R}einforced
\underline{D}istillation).
As described in Algorithm \ref{alg:lord}, LoRD follows a reinforcement
learning paradigm, that is, it consists
of several \emph{periods}, and in each period, the model will learn to
explore new responses and attempt to enhance the model
trained in the last period. However, different from LLMs' alignments,
the agent can neither obtain the reward from the reward model
directly, nor label positive and negative responses manually. This
motivates us to design a new RL method which can \emph{implicitly} measure the reward
for generated tokens under the guidance of victim model's responses.

Illustrated by Figure \ref{fig:lord}, LoRD first requires the model to sample two
sentences randomly at period
$t-1$, which are denoted as $\mathbf{y}_{t-1}^{+}$ and
$\mathbf{y}_{t-1}^{-}$, respectively. In a new period $t$, it first computes the changes of likelihoods
for these two sentences, among the old model $P_{\theta_{t-1}}$ and
the current model $P_{\theta_{t}}$. These changes of likelihoods, denoted
as $\Delta_{t}^{+}$ and $\Delta_{t}^{-}$, indicate whether a selected
sentence is locally \emph{isotropic} ($\Delta >0$) to the optimization
direction with victim
model's response $\mathbf{y}_{vic}$ or not ($\Delta \leq 0$), which
can be seen as the feedback signal for $P_{\theta_{t}}$ in the current optimization step. For convenience, we may swap $\mathbf{y}_{t-1}^{+}$ with
$\mathbf{y}_{t-1}^{-}$ to make sure that
$\Delta_{t}^{+}>\Delta_{t}^{-}$ always holds. In this way, for pairs
$(\mathbf{x},\mathbf{y}_{vic})$ we can take $\mathbf{y}_{t-1}^{+}$ as a
\emph{locality neighborhood} of $\mathbf{y}_{vic}$ and $\mathbf{y}_{t-1}^{-}$ as the
negative sample, all of which can be utilized in the training of $P_{\theta_{t}}$. Figure \ref{fig:explain} illustrates this procedure.
Additionally, LoRD takes $\mathbf{y}_{t-1}^{+}$ as the positive label under the current scope only when $\Delta^{+}$ or $P_{\theta_{t}}(\mathbf{y}_{t-1}^{+}|\mathbf{x})$ exceed their respective fixed thresholds $\tau_{1}$ and $\tau_{2}$. If these conditions are not met, it will use $\mathbf{y}_{vic}$ as a substitute for $\mathbf{y}_{t-1}^{+}$ to enable a cold start.

Based on $\mathbf{y}_{vic}$, $\mathbf{y}_{t-1}^{+}$, and
$\mathbf{y}_{t-1}^{-}$, we now design LoRD's loss function.

\subsection{Design of Loss Functions}\label{sec:der}

From Section \ref{sec:pgm}, we know that the loss function of a policy gradient model can be expressed as an \emph{objective function} to maximize the rewards of decisions (see Equation \ref{eq:pg}) and a \emph{regularization term} to ensure the stability of training. Following this paradigm, the loss function of LoRD could be
\begin{equation}
\label{eq:lord-overall}
\mathcal{L}_{\text{LoRD}}=\mathcal{L}_{obj}+\mathcal{L}_{reg}.
\end{equation}

\noindent
\textbf{Objective function $\mathcal{L}_{obj}$}.
Inspired by the reward model $R_{\theta_{\phi}}$ existed in Equation
\ref{eq:align}, which is trained to
distinguish between positive and negative samples, we propose utilizing the logarithmic
proportion of positive to negative samples as the means of achieving a
de-biased reward, i.e.,
\begin{equation}\small
\label{eq:obj}
\begin{aligned}
\mathcal{L}_{obj}&=-\sum_{\mathbf{x}\in \mathcal{D}_{q}}
                   \text{log}
[\frac{P_{\theta_{t}}(\mathbf{y}_{t-1}^{+}|\mathbf{x})}{P_{\theta_{t}}(\mathbf{y}_{t-1}^{-}|\mathbf{x})}]\\&=-\sum_{\mathbf{x}\in \mathcal{D}_{q}} [{\text{log}P_{\theta_{t}}(\mathbf{y}_{t-1}^{+}|\mathbf{x})-\text{log}P_{\theta_{t}}(\mathbf{y}_{t-1}^{-}|\mathbf{x})}].
\end{aligned}
\end{equation}
Equation \ref{eq:obj} exhibits similarities to previous studies on RL-enhanced LLM~\citep{rl-deduct1,rl-deduct2,rl-deduct3,rl-deduct4,dpo}.
We provide a theoretical explanation for its consistency with the
learning procedure of RLHF and the deduction procedure,
as detailed in Section \ref{sec:th-an} and Appendix \ref{sec:proof1}.

However, training the local model merely by $\mathcal{L}_{obj}$ is
ineffective due to two reasons: \emph{i)} when $\mathcal{L}_{\text{
    LoRD}}:=\mathcal{L}_{obj}$, no information from the victim
model's responses is incorporated into the selection of $\mathbf{y}^{+}_{t-1}$
beyond the cold start phase,
resulting in a meaningless \emph{self-reward-based
  learning} loop for the stealing procedure; \emph{ii)} the
convergence of the local model's training cannot be guaranteed.

To address these two issues simultaneously, we design the
regularization term as follows.

\noindent
\textbf{Regularization loss $\mathcal{L}_{reg}$}. Different from
LLM's RLHF~\citep{trpo,dpo,rlhf1} that typically constrain
$\theta_{t}$ with initial model's generating distribution
$P_{\theta_{init}}(\cdot|\mathbf{x})$, LoRD aims to directly constrain
$\theta_{t}$ with victim model's distribution $P_{\theta_{vic}}(\cdot|\mathbf{x})$.

Unfortunately, $P_{\theta_{vic}}(\cdot|\mathbf{x})$ is typically
\textbf{inaccessible} within the APIs of commercial LLMs and is not feasible for our
black-box scenarios.
Consequently, we incorporate the regularization techniques employed in PPO and TRPO but
tailor our regularization as a bounded contrastive term between the
likelihood of $\theta_{t}$ under the victim model's response and the
negative sample, i.e.,
\begin{equation}\label{eq:reg}\small
\begin{aligned}
  \mathcal{L}_{reg}&=-\sum_{\mathbf{x}\in \mathcal{D}_{q}} clip(\text{log}[\frac{P_{\theta_{t}}(\mathbf{y}_{vic}|\mathbf{x})}{P_{\theta_{t}}(\mathbf{y}_{t-1}^{-}|\mathbf{x})}])\\&=-\sum_{\mathbf{x}\in \mathcal{D}_{q}} clip(\text{log}{P_{\theta_{t}}(\mathbf{y}_{vic}|\mathbf{x})}-\text{log}{P_{\theta_{t}}(\mathbf{y}_{t-1}^{-}|\mathbf{x})}).
\end{aligned}
\end{equation}
In Equation \ref{eq:reg}, we utilize PPO's $clip(\cdot)$ function to
limit the value of the regularization term, as we expect the
regularization term could only be used to avoid the \emph{off the
  cliff} problem~\citep{ppo,trpo} in RL's convergence. Besides, our
contrastive term can be seen as a streamlined black-box variant of the KL
divergence in TRPO. This simplification offers two advantages:
\emph{i)} it alleviates the necessity of loading the initial model's
weights, leading to a substantial reduction in GPU memory usage; \emph{ii)}
it eliminates the need for $P_{\theta_{t}}(\cdot|\mathbf{x})$, which
would otherwise necessitate an additional
exponential operation of $\text{log}P_{\theta_{t}}(\cdot|\mathbf{x})$
that slow down the forward
process and increase extra consumption.\footnote{\emph{logsoftmax} is preferred in the
  implementation of deep learning frameworks~\citep{torch}, as the
  exponential operation in \emph{softmax} and the logarithmic
  operation in \emph{cross-entropy} can be canceled out by each
  other.}

Incorporating Equation \ref{eq:obj} with Equation \ref{eq:reg}, we can
reshape the loss function of LoRD as
\begin{equation}
\label{eq:lord2}\small
\begin{aligned}
  \mathcal{L}_{\text{LoRD}}&=\mathcal{L}_{obj}+\mathcal{L}_{reg}\\&=\sum_{\mathbf{x}\in\mathcal{D}_q} \text{log}[\frac{P_{\theta_{t}}(\mathbf{y}_{t-1}^{-}|\mathbf{x})}{P_{\theta_{t}}(\mathbf{y}_{t-1}^{+}|\mathbf{x})}]+clip(\text{log}[\frac{P_{\theta_{t}}(\mathbf{y}_{t-1}^{-}|\mathbf{x})}{P_{\theta_{t}}(\mathbf{y}_{\text{vic}}|\mathbf{x})})].
\end{aligned}
\end{equation}

Finally, we wrap $\mathcal{L}_{\text{LoRD}}$ with a sigmoid function $\sigma(\cdot)$ to normalize the loss to the interval $(0,1)$:
 \begin{equation}
 \label{eq:black}
 \small
 \begin{aligned}
  \mathcal{L}=\sum_{\mathbf{x}\sim \mathcal{D}_{q}}~\sigma(\text{log}[\frac{P_{\theta_{t}}(\mathbf{y}_{t-1}^{-}|\mathbf{x})}{P_{\theta_{t}}(\mathbf{y}_{t-1}^{+}|\mathbf{x})}]+clip(\text{log}[\frac{P_{\theta_{t}}(\mathbf{y}_{t-1}^{-}|\mathbf{x})}{P_{\theta_{t}}(\mathbf{y}_{{vic}}|\mathbf{x})}])).
 \end{aligned}
 \end{equation}


%% file: analysis.tex
\section{Theoretical Analysis}\label{sec:th-an}

This section will compare LoRD with current model extraction methods
from a theoretical perspective. We will first reveal the underlying
inconsistency between the optimization of LLMs, which typically
involves RL-based alignments, and the previous MEAs utilizing
\emph{MLE} and \emph{KD}. Subsequently, we will demonstrate the
reasons why LoRD can achieve stronger watermark resistance and higher
query efficiency than existing methods.

\subsection{Consistency Analysis on Learning Tasks}\label{sec:ana-tc}
Based on the analysis of the four objective functions for MLE, KD, RLHF
and LoRD, we reach the Proposition \ref{th:con}, and illustrate their
convergence procedure exhibited in Figure
\ref{fig:conv-procedure}. A detailed explanation for it can be found in Appendix \ref{sec:proof1}.

\begin{figure}
  \centering
\includegraphics{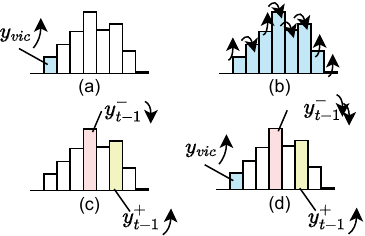}
\caption{\small Illustrations for the converging procedure of
  probability distributions regarding four methods, namely
  MLE (a), KD (b), RLHF (c), and LoRD (d).
  Arrows indicate the expected optimization direction. We mark the distribution dimensions learned with labels in \emph{blue}, and employ \emph{pink} and \emph{yellow} components to indicate the probabilities of positive and negative tokens, respectively.
}
\label{fig:conv-procedure}
\end{figure}
\begin{proposition}[Consistency in Stealing Procedure]\label{th:con}
The learning procedure for LLMs' alignments is consistent with the stealing procedure of LoRD, i.e., they both attempt to maximize the difference between the probabilities of positive and negative samples. Conversely, they are inconsistent with either MLE or KD. In MLE, the objective is maximizing the label probability, while KD aims to minimize the distance among all dimensions.
\end{proposition}

Albeit the inconsistency in their \emph{training procedures}, we put
forward Proposition \ref{th:eqc} to demonstrate that \emph{with enough samples}, all these methods will reach the same distribution results. 

\begin{proposition}[Equivalence when Converged]\label{th:eqc}
  Ideally, for any loss value of Equations \ref{eq:ce}, \ref{eq:r}, \ref{eq:align}, \ref{eq:lord2}, or \ref{eq:black} converging to $0$, we have $\mathbf{y}^{+}\equiv \mathbf{y}_{vic}$. Meanwhile, the local model's distribution $P_{\theta}(\cdot |\mathbf{x})$ will approach that of the victim model $P_{\theta_{vic}}(\cdot |\mathbf{x})$ on MEAs from all three discussed MEA methods, including LoRD, MLE, and KD.
\end{proposition}

Proposition \ref{th:eqc} ensures that the local model will converge to
the victim model \textbf{regardless} of the choice of MEA methods. So
what is the benefit of LoRD? In Section \ref{sec:q-eff}, we will show
that LoRD outperforms current MEAs with two aspects: the query time
reduction, and the
watermark resistance of the learned local model.

\subsection{Comparative Analysis on Model Stealing}\label{sec:q-eff}

\noindent
\textbf{Query Efficiency.} Let $N_{Q}$ and $N_{R}$ denote the sequence
lengths of the query text and the response text, respectively. For
MLE, the \emph{ideal} query numbers to populate the entire text space
are given by $\mathcal{O}(V^{N_{Q}}\cdot V^{N_{R}})$, where $V$
represents the size of the vocabulary. In contrast, LoRD possesses the
capability to automatically explore the generation token space,
thereby significantly reducing the
query requirements about generation candidates to a constant level. Specifically,
the complexity of LoRD's query requirements is
$\mathcal{O}(V^{N_{Q}}\cdot C)$, where
$C$ is a constant that correlates with the capability of local models.

Based on the above analysis, a straightforward concern with employing
MLE in LLMs' extraction is that, given the limited query times in
real-world practices, it may suffer from incomplete learning,
especially for text generation tasks. Consequently, the local model
may tend to memorize some specific responses instead of achieving a
broad understanding and generation. We call such a phenomenon
\emph{preference overfitting (PO)}, which indicates that the local
model is only effective on a limited set of explored samples, and yet
does not generalize well to unseen scenarios. In such cases, the local
model usually exhibits a more ``rugged'' decision surface, which
appears to \emph{overfit} the preference sentences in
$\mathcal{D}_{tr}$, as shown in Figure \ref{fig:intro2} (b). Figure
\ref{fig:distviz} provides a visualization of it.

\noindent
\textbf{Watermark Resistance.}
Another limitation of prevalent objective functions, such as MLE
and KD, is their susceptibility to watermarks~\citep{wm1,wm2,wm4,wm6} of
output contents, i.e., while stealing knowledge from LLMs via
responses $\mathbf{y}_{vic}$, watermarks within them will also been
passively inherited by the local model. Consequently, the generated sentences of the local model may possess some \emph{residual} of watermarks, which might be detected as evidence of stealing.

Despite introducing current watermark removal techniques, we indicate that LoRD can mitigate the influences of watermarks
naturally, as it does not learn the likelihood of victim models' responses $\mathbf{y}_{vic}\sim \mathcal{D}_{tr}$ directly, but relies on
$\mathbf{y}_{vic}$ to determine positive and negative labels
from responses generated by the local model.

\definecolor{lzgreen}{HTML}{468585}
\definecolor{lzblue}{HTML}{2C74B3}
\definecolor{lzred}{HTML}{f78fb3}
\definecolor{lzgrey}{HTML}{bdb5a6}

\newcommand{\red}{\cellcolor{lzred!68} } 
\newcommand{\gret}{\cellcolor{lzred!48} } 
\newcommand{\gres}{\cellcolor{lzred!28} } 
\newcommand{\grem}{\cellcolor{lzgreen!18} } 

\newcommand{\grel}{\cellcolor{lzgreen!38} } 
\newcommand{\grexl}{\cellcolor{lzgreen!68} } 

\newcommand{\greyc}{\cellcolor{lzgrey!28} }

As depicted in Equation \ref{eq:obj}, LoRD guides the
local model to learn the likelihood of $\mathbf{y}_{t-1}^{+}$ instead
of $\mathbf{y}_{vic}$, which means that it will not been influenced
by watermarks contained in $\mathbf{y}_{vic}$ explicitly. However, the
regularization term $\mathcal{L}_{reg}$, as well as the replacement
$\mathbf{y}_{t-1}^{+}\leftarrow \mathbf{y}_{vic}$ for a cold start, will indeed
introduce watermarks from $\mathbf{y}_{vic}$. To address this, we can
reshape Equation \ref{eq:black} into a convex combination of the objective function and the regularization, i.e.,
\begin{equation*}\footnotesize
\label{eq:lambda}
\begin{aligned}
\mathcal{L}&=\mathbb{E} [(1-\lambda_{1})\cdot
(\text{log}P_{\theta_{t}}(\mathbf{y}_{t-1}^{+}|\mathbf{x})-\text{log}P_{\theta_{t}}(\mathbf{y}_{t-1}^{-}|\mathbf{x}))\\&\quad\quad~~+\lambda_{1}\cdot clip(\text{log}P_{\theta_{t}}(\mathbf{y}_{vic}|\mathbf{x})-\text{log}P_{\theta_{t}}(\mathbf{y}_{t-1}^{-}|\mathbf{x}))],
\end{aligned}
\end{equation*}
where $0\leq\lambda_{1}\leq 1$ is the hyperparameter.

When $\lambda_{1}$ is small, the convergence of LoRD will
substantially focus on maximizing
$P_{\theta_{t}}(\mathbf{y}_{t-1}^{+}|\mathbf{x})/P_{\theta_{t}}(\mathbf{y}_{t-1}^{-}|\mathbf{x})$,
with which the local model will exhibit a strong watermark resistance
ability.
When $\lambda_{1}$ increases, LoRD will tend to rely more on the
guidance of $\mathbf{y}_{vic}$, resulting in a higher risk of
introducing watermarks. In the case of $\lambda_{1}=1$, the local
model will converge to the victim model without any exploration and
watermark resistance, which might suffer from the same level of defense by
watermarks.

From a global perspective, $\mathcal{L}_{obj}$ represents the
exploration and the locality learning ability of LoRD, 
which can mitigate the influences of watermarks. On the other hand, $\mathcal{L}_{reg}$ ensures the stability of the training
procedure. Therefore, $\mathcal{L}$ characterizes a
trade-off via $\lambda_{1}$ between the stability and the diversity
during stealing, and Equation \ref{eq:black} can be seen as a special case of $\mathcal{L}$ with $\lambda_{1}=0.5$.

We provide an empirical comparison for query efficiency in Appendix \ref{sec:vary-query},
and the comparison on watermark resistance in
Appendix \ref{sec:eval-wm}.


%% file: exper.tex
\section{Experiments}\label{sec:experiment}

\subsection{Settings}\label{sec:set}

\noindent
\textbf{Datasets.} We evaluate MEAs on six mainstream natural language
generation (NLG) tasks, including \emph{safety alignment}, \emph{machine translation},
\emph{text summarization}, \emph{question answering}, \emph{structured
  text generation}, and  \emph{data-to-text}.
We select twelve representative datasets, including two for safety
alignment and ten for domain-specific evaluation, as detailed in Table
\ref{tab:datas-models}. We believe these datasets encompass the
majority of downstream tasks and effectively capture the varying
degrees of difficulty in model stealing across different task domains.

\noindent
\textbf{Baselines.} As described in Section \ref{sec:bg-lm} and
\ref{sec:ana-tc}, we compare LoRD with two types of model extraction
methods: maximum likelihood estimation (MLE) and knowledge
distillation (KD). For MLE and LoRD, we conduct MEAs under pure
\textbf{black-box} attack settings (see Appendix \ref{sec:threat} for
more details of the threat model). For KD, the predicted distributions are used specifically under grey-box settings.

\noindent
\textbf{Metrics.}
For text generation tasks, we evaluate extracted models with a semantic-level and two lexical-level metrics, BERTScore~\citep{bertscore}, BLEU~\citep{bleu}, and Rouge-L~\citep{rouge}, all of which are commonly used in the NLG evaluation. Regarding reasoning tasks (e.g., QA), we use Precision, Recall, Accuracy, and F1 score as their evaluation metrics.

\noindent
\textbf{Implementation Details.}
We use Llama3-8B as the local model to learn the outputs generated by victim models.
We set sequence length varying 128 to 4096 depending on the selected tasks, and learning rate $3\times 10^{-5}$. Our experiments run on 2 $\times$ 80GB Nvidia Tesla A100. We execute each training five times and record the mean values and standard variances. For LoRD, we set $\tau_{1}$ and $\tau_{2}$ to 0.8 and -0.1, respectively. Besides, we set the period number $N_{t}$ to 512, and use $\lambda_{1}=0.5$.


\renewcommand{\red}{}
\renewcommand{\gret}{}
\begin{table}
\centering
\resizebox{0.49\textwidth}{!}{%
\begin{tabular}{crrrr}
\Xhline{1.5pt}
\multicolumn{1}{l|}{Model/Metric} &
  \multicolumn{1}{c}{BLEU-1} &
  \multicolumn{1}{c}{BLEU-4} &
  \multicolumn{1}{c}{Rouge-L} &
  \multicolumn{1}{c}{BERTScore} \\ \hline
\multicolumn{5}{c}{\emph{Czech to English with 16 query samples}}                                                \\ \hline
\multicolumn{1}{r|}{Victim Model} &\greyc 0.611           & \greyc 0.313           & \greyc 0.604           & \greyc 0.957           \\
\multicolumn{1}{r|}{Local Model}  &\red 0.255         & \red 0.105           & \red 0.348           & \gret 0.868           \\ \hline
\multicolumn{1}{r|}{+MLE}         & \gret $0.535\pm 0.01$ & \gret $0.245\pm 0.01$ & \gret $0.526\pm 0.01$ & \gret $0.899\pm 0.00$ \\
\multicolumn{1}{r|}{+LoRD}        & \gret $0.545\pm 0.01$ & \gret $0.249\pm 0.00$ & \gret $0.538\pm 0.01$ & \gret $0.906\pm 0.00$ \\ \hline
\multicolumn{5}{c}{\emph{German to English with 16 query sample}}                                               \\ \hline
\multicolumn{1}{r|}{Victim Model} & \greyc 0.661           & \greyc 0.377           & \greyc 0.652           & \greyc 0.965           \\
\multicolumn{1}{r|}{Local Model}  & \red 0.276                &\red 0.130                 & \red 0.359 & \gret 0.877               \\
\multicolumn{1}{r|}{+MLE}         & \gret $0.578\pm 0.02$ & \gret $0.302\pm 0.01$ & \gret $0.573\pm 0.02$ & \gret $0.904\pm 0.01$ \\
\multicolumn{1}{r|}{+LoRD}        & \gret $0.587\pm 0.00$ & \gret $0.308\pm 0.00$ & \gret $0.589\pm 0.00$ & \gret $0.917\pm 0.00$ \\ \hline
\multicolumn{5}{c}{\emph{Finnish to English with 16 query samples}}                                              \\ \hline
\multicolumn{1}{r|}{Victim Model} & \greyc 0.558           & \greyc 0.252           & \greyc 0.557           & \greyc 0.953           \\
\multicolumn{1}{r|}{Local Model}  & \red 0.242               & \red 0.085     & \red 0.320 & \gret 0.866                \\
\multicolumn{1}{r|}{+MLE}         & \red $0.444\pm 0.03$ & \gret $0.173\pm 0.02$ & \red $0.449\pm 0.03$ & \gret $0.905\pm 0.00$ \\
\multicolumn{1}{r|}{+LoRD}        & \gret $0.498\pm 0.01$ & \gret $0.196\pm 0.00$ & \gret $0.485\pm 0.01$ & \gret $0.905\pm 0.00$\\
\Xhline{1.5pt}
\end{tabular}%
}
\caption{MEA comparison on WMT16~\citep{wmt16} among MLE and our LoRD methods,
  where we use GPT-3.5-turbo as the victim model, and Llama3-8B~\citep{llama3} as the
local initial model.
}
\label{tab:wmt-res}
\end{table}

\subsection{Stealing Domain-Specific Knowledge}\label{sec:steal-exp}

We first select GPT-3.5-turbo, a checkpoint of ChatGPT, as the basic victim
model. This is because its API provides \emph{probabilities} of
candidate words when generating responses. We employ Llama3-8B~\citep{llama3}, a small LLM
with only a 4.5\% fraction of parameters than the victim model as our
initial local model. Though this LaViSH (\textbf{La}rge-\textbf{Vi}ctim-\textbf{S}mall-\textbf{H}eist) setting contradicts previous
assumptions~\citep{mea-tra-steal,mea-tra-prac,mea-tra-high} in MEA that the copy model should usually be
``wider'' or ``larger'' than the victim model to contain its
knowledge, we believe this setting is more applicable in real world
scenarios~\citep{mea-code}. Appendix \ref{sec:threat} provides more
detail for this setting. Besides, the number of query times selected in this section is
less than 100, a significant degradation compared to previous
studies~\citep{mea-code}. This is because, in our experiments, copy models
can easily learn the knowledge with a few training samples
and then exhibit only slight improvements afterward. More discussions on
query times can be found in Appendix \ref{sec:vary-query}.

\begin{table}[t]
\centering
\resizebox{0.99\linewidth}{!}{%
\begin{tabular}{crrrr}
\Xhline{1.5pt}
\multicolumn{1}{l|}{Model/Metric} &
  \multicolumn{1}{c}{Accuracy} &
  \multicolumn{1}{c}{Precision} &
  \multicolumn{1}{c}{Recall} &
  \multicolumn{1}{c}{F1 Score} \\ \hline
\multicolumn{5}{c}{\emph{PIQA~\citep{piqa} with 64 query samples}}   \\ \hline
  \multicolumn{1}{l|}{Victim Model} &\greyc 0.828& \greyc 0.828 & \greyc 0.827 & \greyc 0.827 \\
\multicolumn{1}{l|}{Local Model}   &\red 0.622  & \red 0.638 & \red 0.621 & \red 0.609\\ \hline
\multicolumn{1}{l|}{+MLE (baseline)}  &\gret $0.760\pm 0.02$  & \gret $0.771\pm 0.01$ & \gret $0.760\pm 0.02$ & \gret $0.757\pm 0.03$\\
\multicolumn{1}{l|}{+KD (gre-box)} & \greyc $0.759\pm 0.02$ & \greyc $0.760\pm 0.02$ & \greyc $0.759\pm 0.02$ & \greyc $0.759\pm 0.02$ \\
\multicolumn{1}{l|}{+LoRD (ours)}  &\gres $0.785\pm 0.01$   & \grem $0.795\pm 0.01$  & \gres $0.785\pm 0.01$  & \gres $0.783\pm 0.02$  \\ \hline
\multicolumn{5}{c}{\emph{TruthfulQA~\citep{tqa} with 64 query samples}} \\ \hline
\multicolumn{1}{l|}{Victim Model} &\greyc 0.414  & \greyc 0.500 & \greyc 0.207 & \greyc 0.293  \\
\multicolumn{1}{l|}{Local Model}  &\grem 0.391  & \grel 0.500 &\grem 0.195  &\grem 0.281  \\
\multicolumn{1}{l|}{+MLE (baseline)}    &\gres $0.381\pm 0.17$ & \grem $0.500\pm 0.00$ & \gres $0.190\pm 0.09$ &\gres $0.266\pm 0.09$  \\
\multicolumn{1}{l|}{+KD (gre-box)} & \greyc $0.463\pm 0.03$ & \greyc $0.500\pm 0.00$ & \greyc $0.232\pm 0.01$ & \greyc $0.316\pm 0.01$ \\
\multicolumn{1}{l|}{+LoRD (ours)}   &\grem $0.408\pm 0.05$  & \grel $0.500\pm 0.00$ &\grel $0.204\pm 0.03$  &\grel $0.289\pm 0.03$  \\
\Xhline{1.5pt}
\end{tabular}%
}
\caption{MEA comparison on QA tasks among MLE and our LoRD methods.
More experiments are shown in Table \ref{tab:domain-res}.}
\label{tab:qa-res}
\end{table}

\noindent
\textbf{Fidelity and limits on stealing.}
We first examine the fidelity and limits of a small LLM to steal commercial
LLMs. As shown in Table \ref{tab:wmt-res}, \ref{tab:qa-res} and \ref{tab:domain-res}, we list the performance
of the victim model and the local model on five tasks, and provide
two MEA methods, local model fine-tuned with MLE (+MLE) and LoRD
(+LoRD), respectively.

We can see that the original performance of the
local model is significantly lower than the victim model, i.e., with a 50\%
decrease in BLEU-4 or $10\sim 25$ decrease in Rouge-L. Once
we employ MEAs in the local model, its performance rapidly boosts to
nearly the same as the victim model, with
$0\sim 40\%$ points of gaps in BERTScore. These gaps are negligible (e.g. $<1\%$ in summarization) in
some tasks, but remain eminent in other tasks such
as reasoning, structured text generation, and machine
translation. This phenomenon indicates that domain-specific model extractions can
effectively learn domain-specific abilities from victim models but may
perform poorly if downstream tasks require extra knowledge, such as
machine translation and QA. We provide a stealing comparison among
different local models in Table \ref{fig:fidelity}.

\noindent
\textbf{Comparison among stealing methods.} Tables
\ref{tab:wmt-res}, \ref{tab:qa-res}, and \ref{tab:domain-res} compare the stealing efficacy between MLE and our
LoRD. The results consistently show that
LoRD outperforms MLE under the same MEA
settings. Besides, for challenging tasks such as reasoning
and translation, LoRD exhibits much higher improvements, which
demonstrates that it can address the preference overfitting problem
discussed in Section \ref{sec:q-eff} and do enable the local model to learn
the task ability from victim models. However, we also observe that for some
tasks (e.g., summarization), LoRD shows no statistical difference from
MLE, probably because these tasks are relatively simple, where merely MLE has
already achieved comparable results to victim models.

\begin{table*}[t]
\centering
\resizebox{0.9\textwidth}{!}{%
\begin{tabular}{l|llllllllll}
\Xhline{1.5pt}
 & \multicolumn{5}{c}{\textbf{DiaSafety}} & \multicolumn{5}{c}{\textbf{SafeRLHF}} \\ \hline
\multicolumn{1}{c|}{Model} & \multicolumn{1}{c}{Toxicity} & \multicolumn{1}{c}{Insult} & \multicolumn{1}{c}{Profanity} & \multicolumn{1}{c}{Severe Toxity} & \multicolumn{1}{c}{Threat} & \multicolumn{1}{c}{Toxicity} & \multicolumn{1}{c}{Insult} & \multicolumn{1}{c}{Profanity} & \multicolumn{1}{c}{Severe Toxity} & \multicolumn{1}{c}{Threat} \\ \hline
Llama3-8B (initial) & 14..20 & 7.94 & 8.35 & 1.58 & 2.29 & 7.92 & 2.71 & 2.80 & 0.30 & 1.49 \\
+MLE & 8.31 & 3.69 & 4.31 & 0.83 & 1.50 & 4.87 & 1.98 & \textbf{1.66} & \textbf{0.16} & 1.02 \\
+LoRD & \textbf{6.45} & \textbf{2.81} & \textbf{3.56} & \textbf{0.71} & \textbf{1.34} & \textbf{3.55} & \textbf{1.15} & 2.84 & 0.38 & \textbf{0.79} \\ \Xhline{1.35pt}
\end{tabular}%
}
\caption{Comparison on safety alignment extraction tasks.}
\label{tab:steal-align}
\end{table*}

\begin{figure}
  \centering
\includegraphics[width=0.95\linewidth]{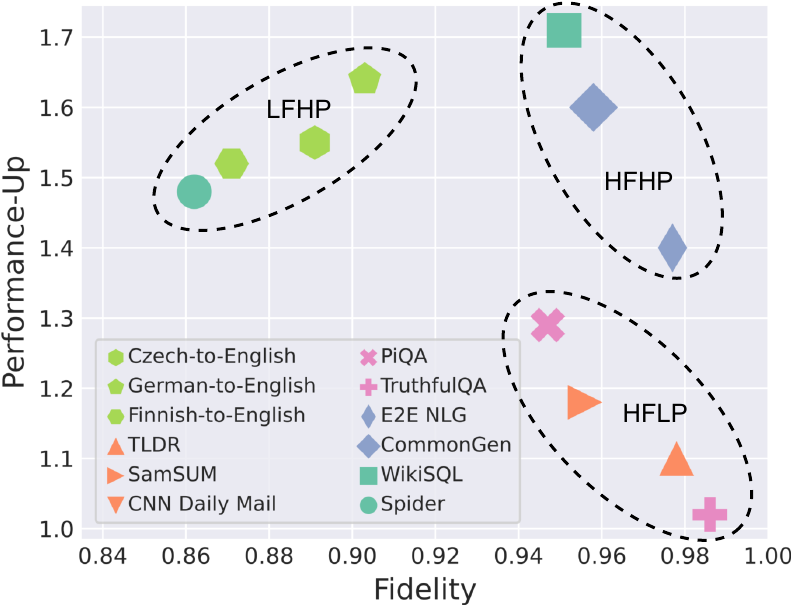}
\caption{Spectrum of the fidelity and performance-up on extracting
  different downstream tasks.
}
\label{fig:spectrum}
\end{figure}

\noindent
\textbf{Tasks difficulties comparison.} Based on previous analysis, we
observe that the performance and limitations of MEA depend on
the category of tasks. Additionally, sometimes datasets in the same task exhibit
significant differences in stealing. We
put forward two metrics to measure task difficulties: the \emph{fidelity} that measures
extraction efficacy compared to victim models, and the
\emph{performance-up}, which assesses the performance gain before
and after stealing for a given local model.
Formally, given a test
set $\mathcal{D}_{te}=\{(\mathbf{x},\mathbf{y})\}$ and a corresponding
metric $\mathcal{M}(hypothesis,reference)$, the fidelity ($F$) and performance-up ($P$) of the
local model $\theta_{N_{t}}$ can be defined as:
\begin{equation}\small
\label{eq:f}
F=\frac{\sum\limits_{\mathbf{x},\mathbf{y}\in
    \mathcal{D}_{te}}\mathcal{M}(\mathbf{y}_{N_{t}},\mathbf{y})}{\sum\limits_{\mathbf{x},\mathbf{y}\in
    \mathcal{D}_{te}}\mathcal{M}(\mathbf{y}_{vic},\mathbf{y})},  P=\frac{\sum\limits_{\mathbf{x},\mathbf{y}\in \mathcal{D}_{te}}\mathcal{M}(\mathbf{y}_{N_t},\mathbf{y})}{\sum\limits_{\mathbf{x},\mathbf{y}\in \mathcal{D}_{te}}\mathcal{M}(\mathbf{y}_{0},\mathbf{y})},
\end{equation}
where $\mathbf{y}_{N_{t}}\sim P_{\theta_{N_{t}}}(\cdot |\mathbf{x})$,
$\mathbf{y}_{0}\sim P_{\theta_{0}}(\cdot |\mathbf{x})$, and
$\mathbf{y}_{vic}\sim P_{\theta_{vic}}(\cdot |\mathbf{x})$ denote the
sampled responses from the trained local model ($\theta_{N_{t}}$), the initial
local model ($\theta_{0}$), and the victim model ($\theta_{vic}$),
respectively. In Figure \ref{fig:spectrum}, we illustrate a ``spectrum'' of extracting
various downstream tasks based on these two metrics defined in Equation
\ref{eq:f}. The figure can assist in recognizing and
defending commercial LLM's knowledge.
From Figure \ref{fig:spectrum}, we observe five tasks
forming the following three scenario groups and datasets coming from
the same tasks are mostly in the same group:

\noindent
$\bullet$ \textbf{High fidelity \& performance-up (HFHP)}. These
  tasks are challenging for a pre-trained model but can be effectively
  learned with the guidance of victim models. This group includes two tasks: data-to-text and structured text generation.

\noindent
$\bullet$ \textbf{High fidelity \& low performance-up (HFLP)}. The
  initial local model already achieves a comparable performance to
  the victim model. QAs and summarization are in this group.

\noindent
$\bullet\,$\textbf{Low fidelity \& high performance-up (LFHP)}. While  MEAs significantly improve the local model's performance, gaps between the local and victim models remain difficult to bridge with domain-specific extraction alone. Machine translation is a representative task whose reasons are explained in Section \ref{sec:steal-exp}.

\subsection{Stealing Safety Alignments}

Besides of the domain-specific model extraction, we also propose the
safety alignment extraction. Specifically, we select two popular
safety alignment datasets for the experiments, namely
SafeRLHF~\citep{saferlhf} and DiaSafety~\citep{diasafety}, to assess the
safety of the generated responses. We employed PerspectiveAPI~\footnote{https://perspectiveapi.com/} to automatically evaluate the
safety of the responses. We select five key aspects of safety
probabilities: Toxicity, Insult, Profanity, Severe Toxicity, and
Threat. In these categories, a lower score indicates better safety
performance. For the LoRD model, we have retained the same
hyper-parameters as those used in our domain-specific experiments to
ensure consistency.
As shown in Table \ref{tab:steal-align}, we can see that both MLE and
LoRD significantly reduce the harmful information after the stealing
procedure. However, LoRD consistantly outperforms MLE on most of the
indicators, suggesting that it can achieve better performance in the
alignment task.


%% file: appendix.tex
\clearpage
\newpage

\input{exper_analysis}

\section{Theoretical Explanations and Proofs}

\subsection{Explanation of Proposition \ref{th:con}}\label{sec:proof1}


As we described in Section \ref{sec:bg}, both existing methods and LoRD are learned from the victim model's response $\mathbf{y}_{vic}$ and the corresponding probability distribution $P_{\theta_{vic}}(\cdot |\mathbf{x}) \in \mathbb{R}^{V}$, where $V$ denotes the vocabulary size. Therefore, we first investigate how the local model is learned to emulate the distribution of the victim model,$P_{\theta_{vic}}(\cdot|\mathbf{x})$, under the following three stealing strategies.

\noindent
\textbf{Expected Distribution of MLE.} We can first reshape the MLE loss into a special formation of Kullback-Leibler divergence with labels of one-hot distributions, that is,
\begin{equation}\label{eq:kl-onehot}\small
  \begin{aligned}
\mathcal{L}_{ce}&=-
                  \sum_{\mathbf{x},\mathbf{y}\sim \mathcal{D}_{tr}}{
                  \text{log}P_{\theta}(\mathbf{y}_{vic}|\mathbf{x})}\\
                &=\sum_{\mathbf{x},\mathbf{y}\sim \mathcal{D}_{tr}}\sum_{j}^{N}{
                  \mathbb{D}_{KL}[\mathbf{1}_{y_{vic,j}}||P_{\theta}(\cdot|\mathbf{x},\mathbf{y}_{vic,<j})]},
  \end{aligned}
\end{equation}
where $\mathbf{1}_{y_{vic,j}}$ is a one-hot vector in which only $\mathbf{1}_{y_{vic,j}}[y_{vic,j}]=1$ and all the other elements are 0. Equation \ref{eq:kl-onehot} demonstrates that MLE learns to maximize the probability of $\mathbf{y}_{vic,j}$, without explicit constraints on probabilities across other dimensions.

\noindent
\textbf{Expected Distribution of KD.}
Following a previous work~\citep{kd}, the objective function of KD is
\begin{equation}\small
\label{eq:kd}
\begin{aligned}
   \mathcal{L}_{kd}&=\mathbb{D}_{KL}[P_{\theta_{vic}}(\cdot |\mathbf{x})||P_{\theta}(\cdot |\mathbf{x})]\\&+T^{2}\cdot \mathbb{D}_{KL}[\text{SM}(P_{\theta_{vic}}(\cdot |\mathbf{x})/T)||\text{SM}(P_{\theta}(\cdot |\mathbf{x})/T)],
\end{aligned}
\end{equation}
where $\text{SM}(\cdot)$ represents the \emph{softmax function}, and
$T>1$ denotes the temperature to smooth the targeted distribution
$P_{\theta_{vic}}(\cdot | \mathbf{x})$. As described in Equation
\ref{eq:kd}, knowledge distillation aims to align $P_{\theta}(\cdot
|\mathbf{x})$ with $P_{\theta_{vic}}(\cdot |\mathbf{x})$ in both the
original and the smoothed probability across all dimensions, which is
exceptionally comprehensive among these MEA loss functions.

\noindent
\textbf{Expected Distribution of Alignments.}
Replacing Equation \ref{eq:align} with Equation \ref{eq:r}, we can merge the optimization target of LLMs' alignments as
\begin{equation}\label{eq:rlog}
\scriptsize
\begin{aligned}
   & \min\limits_{\theta*}  -\sum_{(\mathbf{x},\mathbf{y}^{+},\mathbf{y}^{-}) \sim
\mathcal{D}^{pref}}{\sigma\left(
\text{log}\frac{P_{\theta*}(\mathbf{y}^+|\mathbf{x}) /P_{\theta*}(\mathbf{y}^-|\mathbf{x})}{P_{\theta_{init}}(\mathbf{y}^+|\mathbf{x})/P_{\theta_{init}}(\mathbf{y}^-|\mathbf{x})}\right)} \\
&\Rightarrow \max\limits_{\theta*} \sum_{(\mathbf{x},\mathbf{y}^{+},\mathbf{y}^{-}) \sim
\mathcal{D}^{pref}}{
\text{log} P_{\theta*}(\mathbf{y}^+|\mathbf{x}) - \text{log}
  P_{\theta*}(\mathbf{y}^-|\mathbf{x}),
}
\end{aligned}
\end{equation}
where $\theta *$ denotes the expected parameters of the models as 
\begin{equation}\label{eq:pstar}
    \begin{aligned}
    P_{\theta*}(\mathbf{y}|\mathbf{x})=\frac{1}{Z(x)}P_{\theta_{init}}(\mathbf{y}|\mathbf{x})\cdot e^{\frac{1}{\beta}R_{\phi}(\mathbf{x},\mathbf{y})}.
    \end{aligned}
\end{equation}

We provide a detailed derivation for Equation \ref{eq:pstar} in
Appendix \ref{sec:th1}. By replacing Equation \ref{eq:rlog} with Equation \ref{eq:pstar}, the expected distribution can be represented as $\mathbf{r}_{i,j}\cdot P_{\theta_{init}}(\cdot | \mathbf{x})$, in which $\mathbf{r}_{i,j}$ indicates the wrapped distribution gain. This distortion aims to maximize the ratio $ P_{\theta} (y^{+}_{j}|\mathbf{x},\mathbf{y}^{+}_{<j}) / P_{\theta} (y^{-}_{j}|\mathbf{x},\mathbf{y}^{-}_{<j})$, and leave the probabilities in other dimensions unconstrained directly.

\noindent
\textbf{Expected Distribution of LoRD.}
Similar to alignments, the expected converging procedure by the objective function $\mathcal{L}_{obj}$ is also intended to maximize the ratio between positive samples and negative samples, i.e., ${P_{\theta_{t}}(\mathbf{y}_{t-1}^{+}|\mathbf{x})}/{P_{\theta_{t}}(\mathbf{y}_{t-1}^{-}|\mathbf{x})}$. Meanwhile, the regularization term ${P_{\theta_{t}}(\mathbf{y}_{vic}|\mathbf{x})}/{P_{\theta_{t}}(\mathbf{y}_{t-1}^{-}|\mathbf{x})}$ will guide the models to maximize the ratio between
$\mathbf{y}_{vic}$ and $\mathbf{y}_{t-1}^{-}$. As the ``standard response'' to be learned, $\mathbf{y}_{vic}$ can be viewed sufficiently as a positive example. Therefore, we can derive that the optimization target of LoRD is consistent with RLHF's optimization, i.e., both encourage local models to maximize the probability proportion between positive and negative samples.

Similar to Equation \ref{eq:pstar} in which the optimized model can be seen as the distortion of the original model $P_{\theta_{init}}$, in LoRD the optimized model can be regarded as the distortion of the local model $P_{\theta_{0}}$, with $P_{\theta_{t}}(\cdot|\mathbf{x})=\mathbf{r}_{i,j}^{t}P_{\theta_{t-1}}(\cdot|\mathbf{x})$ at each step $t$, where the distortion term $\mathbf{r}_{i,j}^{t}$ is intended to jointly maximize $P_{\theta_{t}}(\mathbf{y}_{t-1}^{+}|\mathbf{x})/P_{\theta_{t}}(\mathbf{y}_{t-1}^{-}|\mathbf{x})$ and $P_{\theta_{t}}(\mathbf{y}_{vic}|\mathbf{x})/P_{\theta_{t}}(\mathbf{y}_{t-1}^{-}|\mathbf{x})$, while leaving the probabilities in other dimensions unconstrained directly.

\subsection{The Deduction of Equation \ref{eq:pstar} in Proposition \ref{th:con}}\label{sec:th1}

From Equation \ref{eq:align}, we can get that
\begin{equation*}\small
\label{eq:der1}
\begin{aligned}
&\max\limits_{\theta}{\sum_{\mathbf{x} \sim \mathcal{D}_{q}}{R_{\theta_{\phi}}(\mathbf{x},\hat{\mathbf{y}})}}-\beta\mathbb{D}_{KL}[P_{\theta}(\mathbf{y}|\mathbf{x})||P_{\theta_{init}}(\mathbf{y}|\mathbf{x})]\\
&\Rightarrow \max\limits_{\theta}{\sum_{\mathbf{x} \sim
  \mathcal{D}_{q}}\sum_{\mathbf{y}\sim P_{\theta}(\cdot
  |\mathbf{x})}{R_{\theta_{\phi}}(\mathbf{x},\mathbf{y})}}\\&\quad\quad\quad-\beta [
  \text{log}P_{\theta}(\mathbf{y}|\mathbf{x})-\text{log}P_{\theta_{init}}(\mathbf{y}|\mathbf{x})]\\
&\Rightarrow \min\limits_{\theta}{\sum_{\mathbf{x} \sim
  \mathcal{D}_{q}}\sum_{\mathbf{y}\sim P_{\theta}(\cdot
  |\mathbf{x})}{-\frac{1}{\beta}R_{\theta_{\phi}}(\mathbf{x},\mathbf{y})}}+
  \text{log}\frac{P_{\theta}(\mathbf{y}|\mathbf{x})}{P_{\theta_{init}}(\mathbf{y}|\mathbf{x})}\\
&\Rightarrow \min\limits_{\theta}\sum_{\mathbf{x} \sim\mathcal{D}_{q}}\sum_{\mathbf{y}\sim P_{\theta}(\cdot  |\mathbf{x})}{-\text{log}(\text{exp}(\frac{1}{\beta}R_{\theta_{\phi}}(\mathbf{x},\mathbf{y})))}\\&\quad\quad\quad +
  \text{log}\frac{P_{\theta}(\mathbf{y}|\mathbf{x})}{P_{\theta_{init}}(\mathbf{y}|\mathbf{x})}\\
&\Rightarrow \min\limits_{\theta}{\sum_{\mathbf{x} \sim
  \mathcal{D}_{q}}\sum_{\mathbf{y}\sim P_{\theta}(\cdot
  |\mathbf{x})}{
  \text{log}\frac{P_{\theta}(\mathbf{y}|\mathbf{x})}{\text{exp}(\frac{1}{\beta}R_{\theta_{\phi}}(\mathbf{x},\mathbf{y}))\cdot
  P_{\theta_{init}}(\mathbf{y}|\mathbf{x})}}}.
\end{aligned}
\end{equation*}

If we define a partition function $Z(\mathbf{x})$ with the formation of 

\begin{equation}
\label{eq:zx}
Z(\mathbf{x})=\sum_{\mathbf{y}}{P_{init}(\mathbf{y}|\mathbf{x})\text{exp}(\frac{1}{\beta}R_{\theta_{\phi}}(\mathbf{x},\mathbf{y}))},
\end{equation}
we can reformat the optimization target as
\begin{equation*}\small
\label{eq:der2}
\begin{aligned}
&\min\limits_{\theta}{\sum_{\mathbf{x} \sim
  \mathcal{D}_{q}}\sum_{\mathbf{y}\sim P_{\theta}(\cdot
  |\mathbf{x})}{
  \text{log}\frac{P_{\theta}(\mathbf{y}|\mathbf{x})}{\text{exp}(\frac{1}{\beta}R_{\theta_{\phi}}(\mathbf{x},\mathbf{y}))\cdot
  P_{\theta_{init}}(\mathbf{y}|\mathbf{x})}}}\\
&\Rightarrow \min\limits_{\theta}{\sum_{\mathbf{x} \sim
  \mathcal{D}_{q}}\sum_{\mathbf{y}\sim P_{\theta}(\cdot
  |\mathbf{x})}{
  \text{log}\frac{Z(\mathbf{x})\cdot P_{\theta}(\mathbf{y}|\mathbf{x})}{\text{exp}(\frac{1}{\beta}R_{\theta_{\phi}}(\mathbf{x},\mathbf{y}))\cdot
  P_{\theta_{init}}(\mathbf{y}|\mathbf{x})}}}\\
  &\quad \quad \quad \quad \quad \quad \quad \quad \quad \quad -\text{log}Z(\mathbf{x}).
\end{aligned}
\end{equation*}

If we mark
$\frac{1}{Z(\mathbf{x})}\text{exp}(\frac{1}{\beta}R_{\theta_{\phi}}(\mathbf{x},\mathbf{y}))\cdot
P_{\theta_{init}}(\mathbf{y}|\mathbf{x})$ as
$P_{\theta*}(\mathbf{y}|\mathbf{x})$, then we have
\begin{equation*}\small
\label{eq:der3}
\begin{aligned}
&\min\limits_{\theta}{\sum_{\mathbf{x} \sim
  \mathcal{D}_{q}}\sum_{\mathbf{y}\sim P_{\theta}(\cdot
  |\mathbf{x})}{
  \text{log}\frac{Z(\mathbf{x})\cdot P_{\theta}(\mathbf{y}|\mathbf{x})}{\text{exp}(\frac{1}{\beta}R_{\theta_{\phi}}(\mathbf{x},\mathbf{y}))\cdot
  P_{\theta_{init}}(\mathbf{y}|\mathbf{x})}}}\\&\quad\quad\quad-\text{log}Z(\mathbf{x})\\
&\Rightarrow \min\limits_{\theta}{\sum_{\mathbf{x} \sim
  \mathcal{D}_{q}}\sum_{\mathbf{y}\sim P_{\theta}(\cdot
  |\mathbf{x})}{\text{log}\frac{
P_{\theta}(\mathbf{y}|\mathbf{x})}{P_{\theta*}(\mathbf{y}|\mathbf{x})}}}-\text{log}Z(\mathbf{x}).
\end{aligned}
\end{equation*}

Because $Z(\mathbf{x})$ is independent to $\mathbf{y}$, we can deduct that
\begin{equation}\small
\label{eq:der3}
\begin{aligned}
&\min\limits_{\theta}{\sum_{\mathbf{x} \sim
  \mathcal{D}_{q}}\sum_{\mathbf{y}\sim P_{\theta}(\cdot
  |\mathbf{x})}{\text{log}\frac{
P_{\theta}(\mathbf{y}|\mathbf{x})}{P_{\theta*}(\mathbf{y}|\mathbf{x})}}}-\text{log}Z(\mathbf{x})\\
&\Rightarrow \min\limits_{\theta}{\sum_{\mathbf{x} \sim
  \mathcal{D}_{q}}\left[\sum_{\mathbf{y}\sim P_{\theta}(\cdot
  |\mathbf{x})}{\text{log}\frac{
P_{\theta}(\mathbf{y}|\mathbf{x})}{P_{\theta*}(\mathbf{y}|\mathbf{x})}}\right]-\text{log}Z(\mathbf{x})}\\
&\Rightarrow \min\limits_{\theta}{\sum_{\mathbf{x} \sim
  \mathcal{D}_{q}}\mathbb{D}_{KL}[P_{\theta}(\mathbf{y}|\mathbf{x})||P_{\theta*}(\mathbf{y}|\mathbf{x})]-\text{log}Z(\mathbf{x})}.
\end{aligned}
\end{equation}

As we know that $Z(\mathbf{x})$ does not contain $\theta$, the
above optimization target actually minimizes the KL-divergence
between the distribution of $P_{\theta}$ and $P_{\theta*}$,
demonstrating that $\theta*$ is the optimal value of $\theta$ that satisfies

\begin{equation}\small
\label{eq:der4}
P_{\theta*}(\mathbf{y}|\mathbf{x})=\frac{1}{Z(\mathbf{x})}\text{exp}(\frac{1}{\beta}R_{\theta_{\phi}}(\mathbf{x},\mathbf{y}))\cdot
P_{\theta_{init}}(\mathbf{y}|\mathbf{x}).
\end{equation}

Based on equation \ref{eq:der4}, we can see that the optimal distribution
of $\theta$ is built upon $P_{\theta_{init}}$ with a distortion, as we discussed in Section \ref{sec:ana-tc}.

\subsection{The Proof of Proposition \ref{th:eqc}}

\noindent\textbf{Guarantee of MLE.} From Equation \ref{eq:kl-onehot}
we can obtain that when $\mathcal{L}_{ce}$ decreases to 0, the KL divergence between
$P_\theta(\cdot|\mathbf{x})$ and $P_{\theta_{vic}}(\cdot|\mathbf{x})$
decreases to 0, indicating that
$P_\theta(\cdot|\mathbf{x})$ equals to $P_{\theta_{vic}}(\cdot|\mathbf{x})$.

\noindent\textbf{Guarantee of KD.} As we know,
$\mathbb{D}_{KL}(p,q)\geq 0~\forall~p~\text{and}~q$. Therefore,
if $\mathcal{L}_{kd}$ shown in Equation \ref{eq:kd} equals to 0, then both
$\mathbb{D}_{KL}[P_{\theta}(\cdot|\mathbf{x})||P_{\theta_{vic}}(\cdot|\mathbf{x})]$
and
$\mathbb{D}_{KL}[\text{SM}(P_{\theta}(\cdot|\mathbf{x})/T)||\text{SM}(P_{\theta_{vic}}(\cdot|\mathbf{x})/T)]$
equal to 0. For the latter one, we can derive that only when
$P_{\theta}(\cdot|\mathbf{x})$ equals to
$P_{\theta_{vic}}(\cdot|\mathbf{x})$ can this term reduce to
0 based on the property of KL divergence. Integrating the analysis of these two terms, we can obtain that
$\mathcal{L}_{kd}=0$ represents the local model's distribution
converge to that of the victim model.

\begin{table*}[]
\centering
\resizebox{0.78\linewidth}{!}{%
\begin{tabular}{l|l}
\Xhline{1.25pt}
Task   & Instruction \\ \hline
WMT16 & Please translate the sentence from [source language] to English. \\
PiQA \& TruthfulQA  & Please select the correct answer for the ``Question'' of Users. Question: \\
  &  [question] Selection 1: [Selection1] Selection 2:[Selection2]. \\
E2E NLG & Please translate the information to a sentence in natural language. \\
CommonGen & Please generate a sentence based on the words provided by Users.  \\
WikiSQL\& Spider        & Please return to me the SQL sentence based on the
 text (i.e., Question)  \\
  &  and the table information (i.e., Table) provided by the User.    \\
TLDR\& SamSUM & Please **summarize** the content given by the user. \\
CNN Daily Mail & Please **summarize** the content given by the user. \\
\Xhline{1pt}
\end{tabular}%
}
\caption{Instructions used in the different downstream datasets.}
\label{tab:instruction}
\end{table*}

\noindent\textbf{Guarantee of LoRD.}
When $\mathcal{L}$ shown in Equation \ref{eq:black} equals to 0, the proportion of
$P_{\theta_{t}}(\mathbf{y}_{{vic}}|\mathbf{x})/P_{\theta_{t}}(\mathbf{y}_{{t-1}}^{-}|\mathbf{x})$
and
$P_{\theta_{t}}(\mathbf{y}_{{t-1}}^{+}|\mathbf{x})/P_{\theta_{t}}(\mathbf{y}_{{t-1}}^{-}|\mathbf{x})$
should limit to $-\infty$. As we know that \emph{i)} in a distribution
$\sum{P_{\theta_{t}}(\cdot|\mathbf{x})}=1$ and \emph{ii)} $\mathbf{y}^{+}_{t-1}$
is a dynamic positive response generated at each period, we can deduct
that when $\mathcal{L}=0$ there must be
$\mathbf{y}_{vic}=\mathbf{y}_{t-1}^{+}$, i.e.,
$P_{\theta_{t}}(\mathbf{y}_{vic}|\mathbf{x})=P_{\theta_{t}}(\mathbf{y}_{t-1}^{+}|\mathbf{x})=1$
and $P_{\theta_{t}}(\mathbf{y}_{t-1}^{-}|\mathbf{x})=0$.

Note that
this is merely a theoretical limit for LoRD that cannot be reached, because
$\mathbf{y}_{t-1}^{-}$ will not be sampled if its probability is 0,
and $\mathbf{y}_{t-1}^{+}$ usually doesn't exhibit a significant distinction
to $\mathbf{y}_{t-1}^{-}$ when sampling.

\section{Supplemental Related Works}

\subsection{Human-Feedback-Free Alignments}\label{sec:related-rlaif}

There are several alternatives to the standard RLHF approach~\cite{rlaif,isaac}.
\citet{rlaif} propose reinforcement learning with AI feedback (RLAIF) as a
means to diminish the annotation burden associated with the preference
assessments. Besides, there are some approaches, such as direct preference
optimization (DPO)~\citep{dpo}, that conceptualize the language model itself as the reward model and thus consolidate Equation
\ref{eq:r} and Equation \ref{eq:align} into a unified supervised and
preference-based training task. Since they do not
change the primary targets (i.e., maximizing rewards) and optimization
strategies of LLM's alignments, we only consider the standard formation of alignments for simplicity in our theoretical analysis.

\subsection{Language Models Extraction}\label{sec:related-MEA}

\begin{table}[t]
\centering
\resizebox{0.95\linewidth}{!}{%
\begin{tabular}{l|l}
\Xhline{1.5pt}
Datasets \textbackslash{} Models   & Links \\ \hline
SafeRLHF & \url{https://huggingface.co/datasets/PKU-Alignment/PKU-SafeRLHF} \\
DiaSafety & \url{https://huggingface.co/datasets/thu-coai/diasafety} \\
PIQA & \url{https://huggingface.co/datasets/piqa} \\ 
TruthfulQA & \url{https://huggingface.co/datasets/truthful_qa} \\ 
WMT16 & \url{https://huggingface.co/datasets/wmt16} \\ 
E2E NLG & \url{https://huggingface.co/datasets/e2e_nlg} \\ 
CommonGen & \url{https://huggingface.co/datasets/allenai/common_gen} \\ 
WikiSQL & \url{https://huggingface.co/datasets/wikisql} \\
Spider & \url{https://huggingface.co/datasets/spider} \\
TLDR & \url{https://huggingface.co/datasets/UCL-DARK/openai-tldr-filtered} \\
SamSUM & \url{https://huggingface.co/datasets/samsum} \\
CNN Daily Mail & \url{https://huggingface.co/datasets/cnn_dailymail}\\ \hline
Llama3-8B & \url{https://huggingface.co/meta-llama/Meta-Llama-3-8B-Instruct} \\
Llama3-70B & \url{https://huggingface.co/meta-llama/Meta-Llama-3-70B-Instruct} \\
Phi3-3.8B & \url{https://huggingface.co/microsoft/Phi-3-mini-4k-instruct} \\
OPT-6.7B & \url{https://huggingface.co/facebook/opt-6.7b} \\
Qwen2-7B & \url{https://huggingface.co/Qwen/Qwen2-7B-Instruct} \\
MistralV3-7B & \url{https://huggingface.co/mistralai/Mistral-7B-Instruct-v0.3} \\
  \Xhline{1pt}
\end{tabular}%
}
\caption{Datasets and pre-trained model checkpoints used in the
  paper. Specifically, we select twelve representative datasets:
  SafeRLHF~\cite{saferlhf}, DiaSafety~\cite{diasafety}, WMT16~\cite{wmt16}, TLDR~\cite{tldr}, CNN Daily Mail~\cite{cnn}, Samsum~\cite{samsum}, WikiSQL~\cite{wikisql}, Spider~\cite{spider}, E2E-NLG~\cite{e2enlg}, CommonGen~\cite{commongen}, PIQA~\cite{piqa}, and TruthfulQA~\cite{tqa} as benchmarks for our evaluation. These datasets cover most of the downstream tasks in natural language generation.}
\label{tab:datas-models}
\end{table}

Studies to steal language models originated from the natural
language understanding (NLU) models, such as BERT\citep{bert}, and then
evolved to generative language models, especially large language models recently.

\citet{mea-bertapi} highlights early recognition of model extraction
threats in language models. By constructing text inputs with randomly
vocabulary sampling, they successfully extract the weights from BERT-based
APIs. Besides, \citet{mea-vendor} investigate the feasibility of
side-channel model extraction attacks, revealing that by analyzing
extra signals from GPU kernels, one could accurately steal the model architecture
and its parameters. Subsequent
research~\citep{mea-ensemble} has thoroughly investigated the strategy
of ensembling victim models to train a competitor model that surpasses its teachers.

The exploration of generative language model extraction is still in
its infant stage, with only a handful of studies thus
far. \citet{mea-wmt} investigate imitation attacks on natural language
models. By designing monolingual query texts and collecting responses,
they successfully extract the knowledge from a simulated machine
translation model under the black-box settings. This research exhibits
that slight architectural differences will not influence the
extraction between language models. \citet{mea-code} also explores the
potential risks of stealing the code-generation abilities of LLMs into
smaller downstream models. Unlike previous research~\citep{mea-wmt},
this is the first study that selects LLMs as targets. By collecting
large-scale domain-specific samples, they fine-tune a 7-billion local
pre-trained model with them and show the similarity between the victim
and local models in both performances and adversarial
samples. However, these two studies employ the MLE loss (Equation
\ref{eq:mle}) as the MEA method, neither considering whether MLE is
compatible with LLMs's training, especially the alignment procedure
shown in Section \ref{sec:bg-lm}, nor addressing optimizations related
to query efficiency and the watermark resistance. Besides, the scope
of these studies is limited to stealing specific knowledge in a few
downstream domains. At the same time, most of the critical aspects of
LLMs and the required extraction capabilities, such as query numbers
and local model scales, remain unresolved.
Besides, while various other extraction attacks target LLMs
(e.g., prompt extraction~\cite{prompt}), these lie beyond the scope of our
current discussion.

\subsection{Text Watermarks}
In contrast to stealing LLMs, IP protection methods have received considerable attention recently. By sampling a stealthy but representative ``greed word set'' on the vocabulary distribution, these methods~\citep{wm1,wm2,wm4,wm6} can remap the generated words into their synonyms or add the ``watermarked'' token automatically, and thus effectively certify the output. Besides, strategies such as integrating embeddings into the representation as the backdoor~\citep{model-wm1} or manipulating the probabilities with crafted sinusoidal noises~\citep{wm3,wm7} are also proposed. However, these approaches often presume more stringent conditions regarding the victim and the suspected models. This paper will further assess the effectiveness of LoRD and current MEAs in evading these black-box watermarking strategies.

\section{A Detailed Threat Model}
\label{sec:threat}


\noindent
\textbf{Adversary's Objective.} The adversary's objective is to steal the targeted knowledge from LLMs. Specifically, we select machine translation, reasoning, data-to-text, structured text generation, and summarization as the downstream domain-specific tasks. The adversary aims to develop a \emph{query-efficient} MEA algorithm, since the amount of input and generated tokens will be counted as the costs. Additionally, the MEA methods are expected to be \emph{watermark-resistant}, i.e., they are highly desired to reduce the risks of exposure to unauthorized stealing.

\begin{algorithm}[t]
  {
    \small
\caption{LoRD Algorithm}\label{alg:lord}
\begin{algorithmic}[1] 

\INPUT{Query dataset $\mathcal{D}_{q}$, local language model
    $\theta_{init}$, interface of the victim model
    $P_{\theta_{vic}}(\cdot|\cdot)$, train period number $N_{t}$,
    threshold values $\tau_{1}$ and $\tau_{2}$.}
\STATE \textcolor{blue}{\texttt{// Initialization.}}
\STATE $\theta_{0}\leftarrow \theta_{init},
\mathcal{D}_{tr}\leftarrow \emptyset,
\mathcal{D}_{0}^{+}\leftarrow \emptyset,
\mathcal{D}_{0}^{-}\leftarrow \emptyset,
t\leftarrow 0$;
\STATE \textcolor{blue}{\texttt{// Query the victim model.}}
\FOR{$\mathbf{x} \sim \mathcal{D}_{q}$}
    \STATE $\mathbf{y}_{vic} \leftarrow P_{\theta_{vic}}(\cdot|\mathbf{x})$;
    \STATE $\mathcal{D}_{tr}\leftarrow \mathcal{D}_{tr}\cup \{(\mathbf{x},~\mathbf{y}_{vic},~P_{\theta_{vic}}(\mathbf{y}_{vic}|\mathbf{x}))\}$;
\ENDFOR

\STATE \textcolor{blue}{\texttt{// Train local model.}}
\STATE \textcolor{blue}{\texttt{// Initialize the positive and negative datasets.}}
\STATE $\mathcal{D}_{0}^{+}\leftarrow \mathcal{D}_{q}$;
\FOR{($\mathbf{x}, \mathbf{y}_{vic},
{P}_{\theta_{vic}}(\mathbf{y}_{vic}|\mathbf{x})) \sim
\mathcal{D}_{tr}$}
    \STATE $\mathbf{y}_{0}^{-} \sim P_{\theta_{t}}(\cdot | \mathbf{x})$;
    \STATE $\mathcal{D}_{0}^{-}\leftarrow \mathcal{D}_{0}^{-}\cup
    \{(\mathbf{x}, \mathbf{y}_{0}^{-}, P_{\theta_{0}}(\mathbf{y}_{0}^{-}|\mathbf{x}))\}$;
\ENDFOR

\WHILE{$t < N_{t}$}
    \STATE $t\leftarrow t+1$;
    \STATE $\theta_{t}\leftarrow \theta_{t-1}$;
    \FOR{($\mathbf{x}, \mathbf{y}_{vic},
    {P}_{\theta_{vic}}(\mathbf{y}_{vic}|\mathbf{x})) \sim
    \mathcal{D}_{tr}$}
        \STATE $\mathbf{y}_{t}^{+},\mathbf{y}_{t}^{-} \sim P_{\theta_{t}}(\cdot | \mathbf{x})$;
        \STATE $\mathcal{D}_{t}^{+}\leftarrow \mathcal{D}_{t}^{+}\cup
        \{(\mathbf{x}, \mathbf{y}_{t}^{+})\}$;
        \STATE $\mathcal{D}_{t}^{-}\leftarrow \mathcal{D}_{t}^{-}\cup
        \{(\mathbf{x}, \mathbf{y}_{t}^{-})\}$;
    \ENDFOR

    \STATE \textcolor{blue}{\texttt{// Forward.}}
    \FOR
    {$\mathbf{x},\mathbf{y}_{vic},\mathbf{y}_{t-1}^{+},\mathbf{y}_{t-1}^{-}
      \sim (\mathcal{D}_{tr},\mathcal{D}_{t-1}^{+},\mathcal{D}_{t-1}^{-})
      $}

    \STATE $\Delta^{+}\leftarrow \text{log}P_{\theta_{t}}(\mathbf{y}_{t-1}^{+}|x)-\text{log}P_{\theta_{t-1}}(\mathbf{y}_{t-1}^{+}|x)$;
    \STATE $\Delta^{-}\leftarrow \text{log}P_{\theta_{t}}(\mathbf{y}_{t-1}^{-}|x)-\text{log}P_{\theta_{t-1}}(\mathbf{y}_{t-1}^{-}|x)$;
    \IF{ $\Delta^{+}<\Delta^{-}$}
        \STATE swap$(\mathbf{y}_{t-1}^{+},\mathbf{y}_{t-1}^{-})$;
        \STATE swap$(\Delta^{+},\Delta^{-})$;
    \ENDIF
    \IF{ $P_{\theta_{t}}(\mathbf{y}_{t-1}^{+}|x)<\tau_{1}$ \&\& $\Delta^{+}<\tau_{2}$}
        \STATE $\mathbf{y}_{t-1}^{+}\leftarrow \mathbf{y}_{vic}$;
    \ENDIF

    \STATE \textcolor{blue}{\texttt{// Compute loss with Equation \ref{eq:lord2} or \ref{eq:black}.}}


    \STATE $\mathcal{L}\leftarrow\text{log}[\frac{P_{\theta_{t}}(\mathbf{y}_{t-1}^{-}|\mathbf{x})}{P_{\theta_{t}}(\mathbf{y}_{t-1}^{+}|\mathbf{x})}]+clip(\text{log}[\frac{P_{\theta_{t}}(\mathbf{y}_{t-1}^{-}|\mathbf{x})}{P_{\theta_{t}}(\mathbf{y}_{\text{vic}}|\mathbf{x})})]$;

    \STATE \textcolor{blue}{\texttt{// Backward.}}
    \STATE $\theta_{t}\leftarrow \text{stepUpdate}(\theta_{t}, \mathcal{L})$;
    \ENDFOR
\ENDWHILE
\RETURN $\theta_{t}$
\end{algorithmic}
}
\end{algorithm}

\noindent
\textbf{Targeted Models.} We select Llama3-70B, GPT-3.5-turbo,  and
GPT-4o as the victim models in this paper. Unlike previous works that
only deployed simulated local victim models (e.g., OPT~\citep{opt}), our selections aim to expose the stealing threat on realistic AI services. Besides, our target models are specifically constrained to LLMs fine-tuned with alignment methods (e.g., RLHF) since they are not only state-of-the-art solutions now but also more valuable due to their human-based alignments.

\noindent
\textbf{Adversary's Capabilities.} In accordance with the LLM-based AI service APIs, we identify two attack scenarios: black-box and grey-box attacks. In the black-box scenario, only textual responses the adversary is allowed to obtain. At the same time, all other information, such as the temperature, sampling strategies, and the hidden states of LLMs, are unseen and inaccessible. On the contrary, a grey-box attack allows the adversary to access the generation probabilities distribution of tokens. Notice that both MLE and our LoRD method are under black-box settings, and we only adopt grey-box settings on some particular stealing methods, such as knowledge distillation.

Besides, this paper posits that the adversary usually has worse training conditions than the victims. Specifically, query times and the scale of the local model available to the adversary are much smaller than the victims' training datasets and model parameters. This setting has been adopted in previous LLMs' extraction~\citep{mea-code}. We call it a LaViSH (\textbf{La}rge-\textbf{Vi}ctim-\textbf{S}mall-\textbf{H}eist) framework, which allows us to estimate the upper bound of MEA risks empirically. For adversaries with more substantial resources, they can train more powerful MEA-based LLMs by leveraging MEA algorithms under our LaViSH settings.



%% file: exper_analysis.tex
\section{Supplemental Experiments}

\subsection{Resistance to Watermarks}\label{sec:eval-wm}
Current LLM watermarking methods have been shown~\citep{wm6} to
be robust against commonly used erasing strategies (e.g., rephrasing),
making watermark removal a distinct challenge. In this section, we
validate the inherent resistance of LoRD to watermarks, suggesting
that LoRD is preliminarily resistant to text watermarking.
As described in Section \ref{sec:th-an}, we highlight that
LoRD can extract the victim models' knowledge with two terms: the
straightforward likelihood learning term
$\text{log}P_{\theta_{t}}(\mathbf{y}_{vic}|\mathbf{x})-\text{log}P_{\theta_{t}}(\mathbf{y}_{t-1}^{-}|\mathbf{x})$
and the exploration term
$\text{log}P_{\theta_{t}}(\mathbf{y}_{t-1}^{+}|\mathbf{x})-\text{log}P_{\theta_{t}}(\mathbf{y}_{t-1}^{-}|\mathbf{x})$,
where we can tune $\lambda_{1}$ as shown in
$\mathcal{L}$ to trade off the exploration and the
convergence speed. Typically, a lower $\lambda_{1}$ encourages the
model for conducting a slower but more diverse and localized exploration
from its own generated text $\mathbf{y}_{t-1}^{+}$, potentially enhancing watermark resistance. In this subsection, we evaluate this analysis
empirically.

\noindent
\textbf{Watermarking Details.}
Unlike previous experimental settings in
Section \ref{sec:experiment}, here we cannot utilize commercial LLMs as
victim models due to the inability to control token sampling
inside LLMs. Instead, we employ Llama3-70B as the victim model and
watermark its outputs based on \emph{``green'' tokens selection}. Following
prior research~\citep{wm6}, we separate the predicted
vocabulary into a \emph{green word set} and a \emph{red word set}
, assigning them randomly with the seed derived from the hash values of
generated tokens at the last generation step. Subsequently, we sample the next token
\textbf{exclusively} from the green set, determined by a certain probability.

In this way, given the hypothesis $H_{0}$ that \emph{texts are generated
  without the knowledge of the green word set}, we can estimate the
probability $H_{0}$ occurs (\emph{P-value}) and the \emph{Z-score} of it for these texts. A high P-value, among with a low Z-score, indicates stronger watermark resistance for MEA algorithms.

\begin{figure}[t]
  \centering
\includegraphics[width=0.98\linewidth]{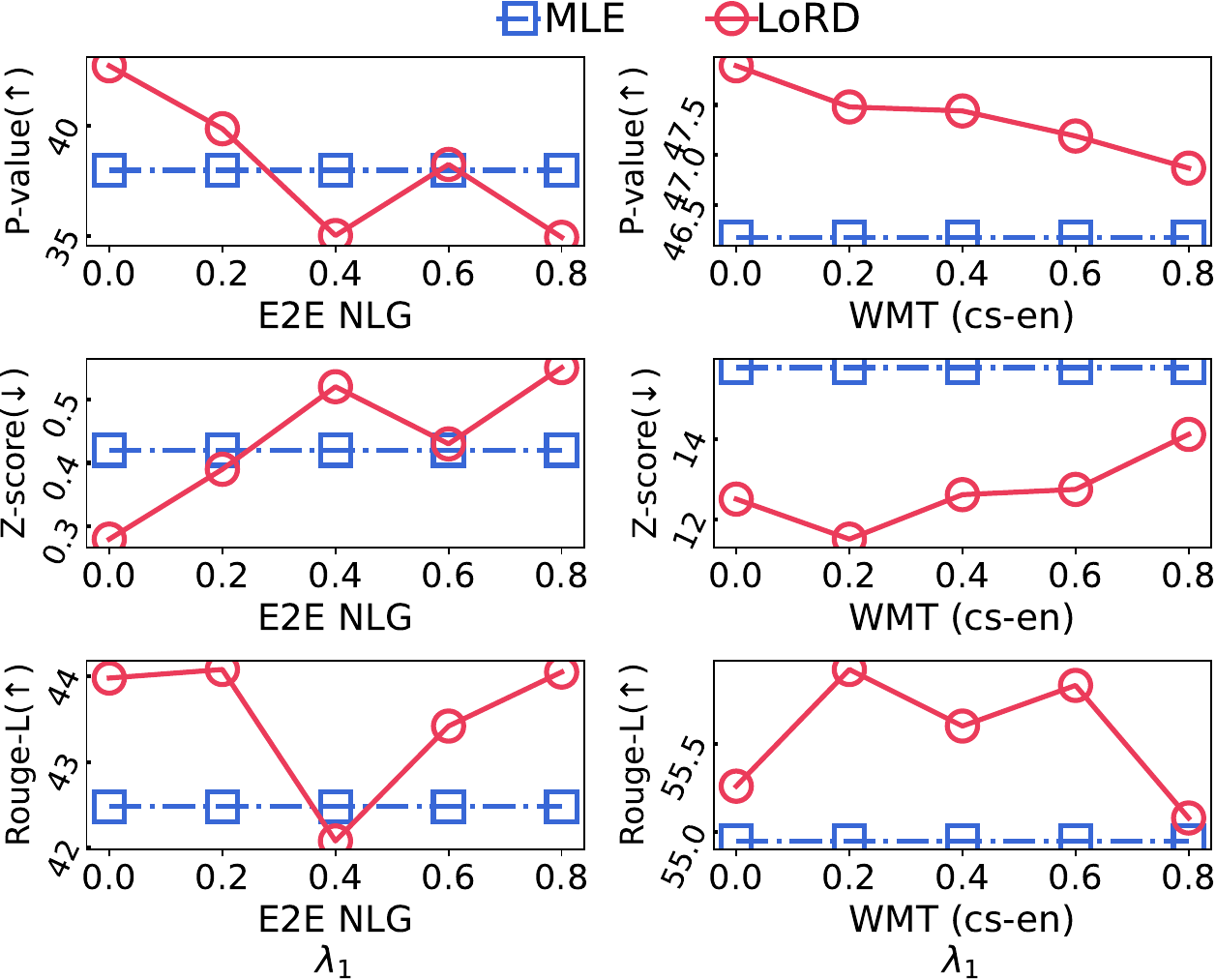}
\caption{Comparison of watermarks resistance.
}
\label{fig:wtmk}
\end{figure}

\noindent
\textbf{Result Analysis.}
As depicted in Figure \ref{fig:wtmk}, we
evaluate the watermark resistance for both MLE and LoRD, and
demonstrate how LoRD's performance varies with different values of
$\lambda_{1}$. The Z-score of
LoRD witnesses a consistent increase as $\lambda1$ arises,
indicating that the ``confidence'' in rejecting the hypothesis, i.e.,
the risk to be suspected, arises when $\lambda_{1}$ increases. This
finding coincides with the analysis in Section \ref{sec:th-an}.
Besides,
we observe that the P-values of LoRD are generally higher than those of MLE when $\lambda_{1}$ is below 0.8, indicating that LoRD typically exhibits stronger watermarking resistance
than MLE in most situations. It is noteworthy that this enhanced
resistance seems not a ``tax'' of MEAs efficacy, as the
Rouge-L (F1) scores of LoRD consistently surpass those of MLE and do not exhibit
a significant negative correlation with their P-values.

\begin{figure*}[t]
  \centering
\includegraphics[width=0.98\linewidth]{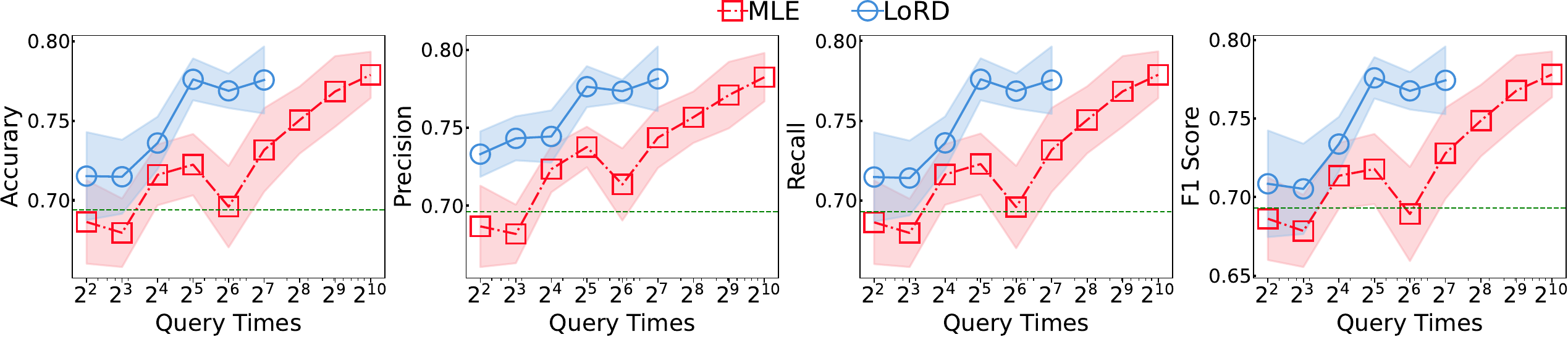}
\caption{Comparison of query efficiency between MLE and LoRD on PiQA, where
  the \emph{green horizontal line} represents the performance of the
  initialized local model. We increase query times for each
  method until reaching their bottlenecks. 
  It can be found that the model extracted by LoRD typically performs a higher accuracy than MLE under the same number of queries. At the same time, LoRD reaches bottlenecks significantly earlier, reducing about 87\% query cost compared with MLE.}
\label{fig:vary-qt}
\end{figure*}

\subsection{Scaling the Stealing}

In this subsection, we explore essential capacities to steal domain-specific
knowledge from LLMs. We first analyze the influence of
query times for the adversary, then compare the efficacy when utilizing
different sizes of the local model, and finally compare the
fidelity among different victim and local models.

\subsubsection{Query Times}\label{sec:vary-query}
We first investigate the influence of query numbers on
MEAs. Specifically, we sample query examples randomly from the query
dataset, starting from 4, and incrementally increase it until the
performance of the learned model stabilizes. Figure
\ref{fig:vary-qt} illustrates the stealing efficacy of LoRD and MLE
on PiQA.

We observe that the scores of MLE and LoRD
consistently increase as the query number rises, showing that a
larger query number can improve stealing efficacy steadily until
reaching their empirical upper bounds. Additionally, LoRD typically obtains a
higher score than MLE with the same number of queries, and reaches
bottlenecks earlier, which can reduce the required query numbers by
87\% compared to MLE. Moreover, in Figure \ref{fig:vary-qt}, the
performance of LoRD exhibits a relatively lower standard variance than
MLE, indicating a more stable training procedure.

\begin{figure*}[t]
\includegraphics[width=1.00\linewidth]{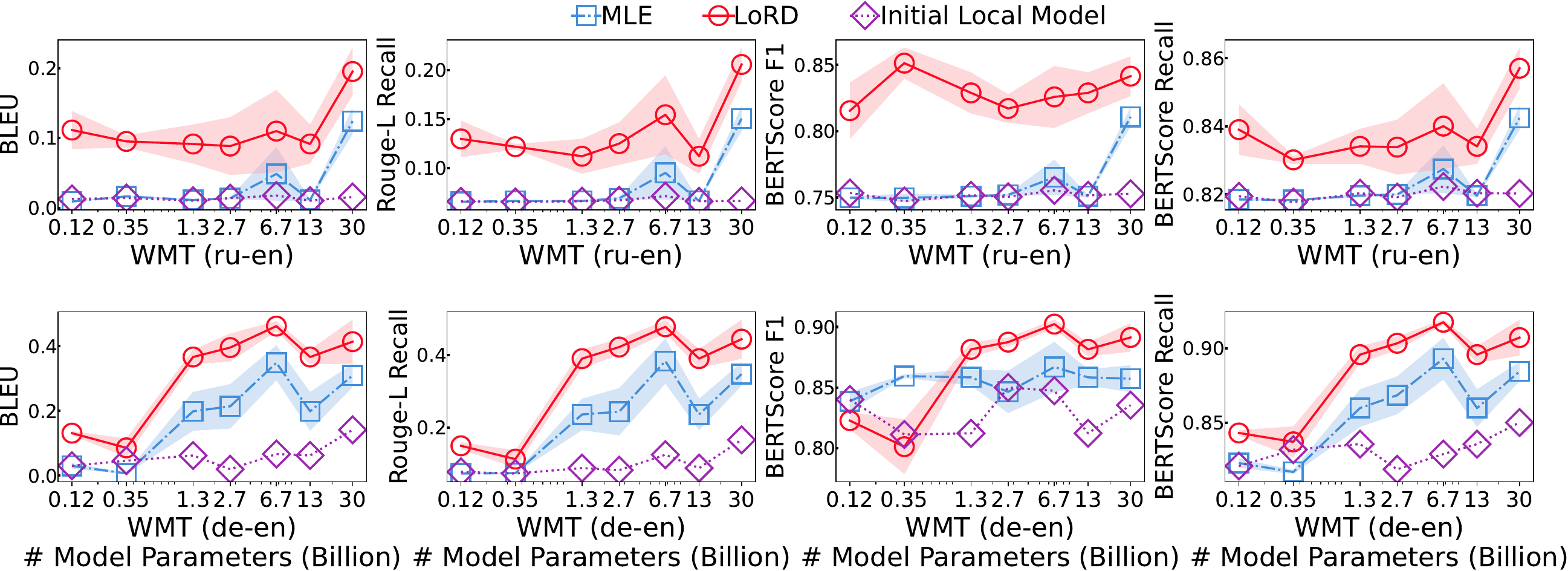}
\caption{Experiments varying different model parameter scales.}
\label{fig:varymodelsize}
\end{figure*}

\subsubsection{Scales of Local Models}\label{sec:scales}
As shown in our threat model (see Appendix \ref{sec:threat}), we assume the adversary is stealing existing commercial LLMs with a small local model. This raises the question of selecting an appropriate interval of the local model's size. To address this concern, we illustrate the correlation between the local model's size and extraction efficacy on two machine translation tasks, Russian-to-English (ru-en) and German-to-English (de-en), as shown in Figure \ref{fig:varymodelsize}. Here, we employ seven OPT models~\citep{opt} as local models, with parameters ranging from 125 million to 30 billion, to minimize the interruptions of factors other than model size.

Figure \ref{fig:varymodelsize} shows a sharp distinction between two machine translation tasks. In the de-en task, the performance of the local model increases steadily with model size, while this trend is not evident in the ru-en task with model size smaller than 30 billion. Nevertheless, the performance of a 30 billion parameter learned local model in ru-en cannot even be comparable to that of a 1.3 billion parameter local model in the de-en task. This phenomenon suggests that for tasks requiring commonsense knowledge, such as machine translation, the local model should at least possess foundational knowledge of the task (e.g., pre-trained on Russian texts) to learn from victim models effectively. Besides, experiments in BERTScore (F1) show that sometimes LoRD may underperform MLE when the local model has fewer than 1 billion parameters, demonstrating that it is challenging to bootstrap LoRD's exploration with a very small local model. By summarizing the increase in LoRD's curves, a model with 2.7 billion appears sufficient to steal domain-specific knowledge from commercial LLMs.

\subsubsection{Fidelity under Different Victim and Local Models}

\begin{figure}
\includegraphics[width=1.00\linewidth]{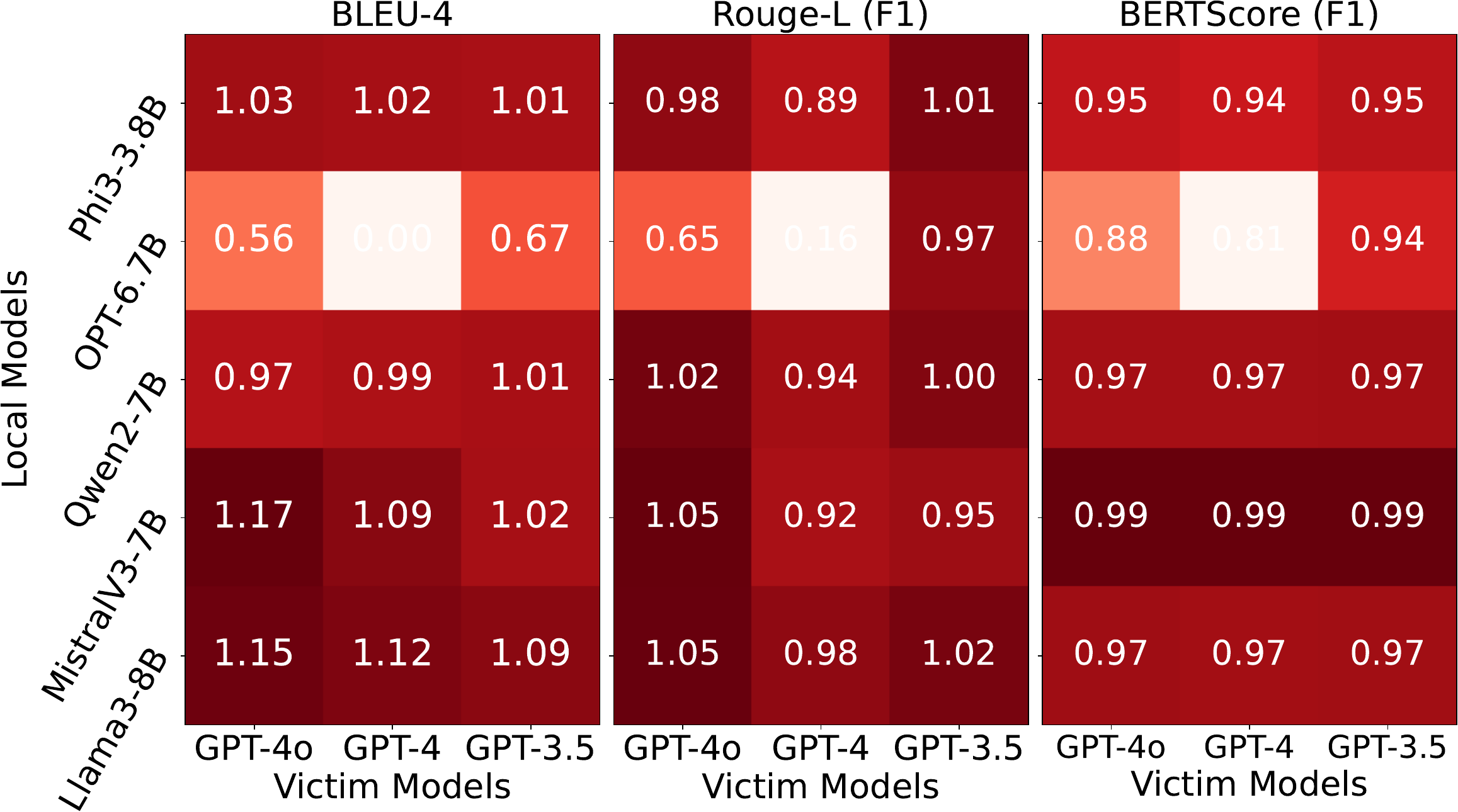}
\caption{Fidelity of extracted models with different victim models
  (GPT-3.5-turbo, GPT-4, and GPT-4o) and different local models
  (Phi-3, OPT, Qwen2, MistralV3, and Llama3).
}
\label{fig:fidelity}
\end{figure}
We 
evaluate the fidelity of extracting different victim models using
various pre-trained local models. Specifically, we select GPT-3.5,
GPT-4, and GPT-4o as victim models, and employ five state-of-the-art
open-source models, Phi-3 (3.8B), OPT (6.7B), Qwen-2 (7B), Mistral-V3
(7B), and Llama-3 (8B), as local models, as shown in Figure \ref{fig:fidelity}.

Horizontally, while GPT-4 exhibits a consistently lower extracted
fidelity compared to the other two victim models, vulnerabilities of
the three victim models are generally similar. Vertically, fidelity
of different local models can be significantly impacted by their
performance. For instance, OPT (6.7B) shows a noticeably lower score
compared to the other four models, which indicates that the initial
performance of the local model will affect the performance of
MEAs. Besides, Phi-3 (3.8B) achieves a comparable
fidelity to larger models like Llama-3 (8B), demonstrating that the size
of a local model does not influence final fidelity in
domain-specific stealing after 2.7 billion, which corroborates the
observation in Appendix \ref{sec:scales}.



\subsection{Visualization of Distributions}

\begin{figure}
\includegraphics[width=0.99\linewidth]{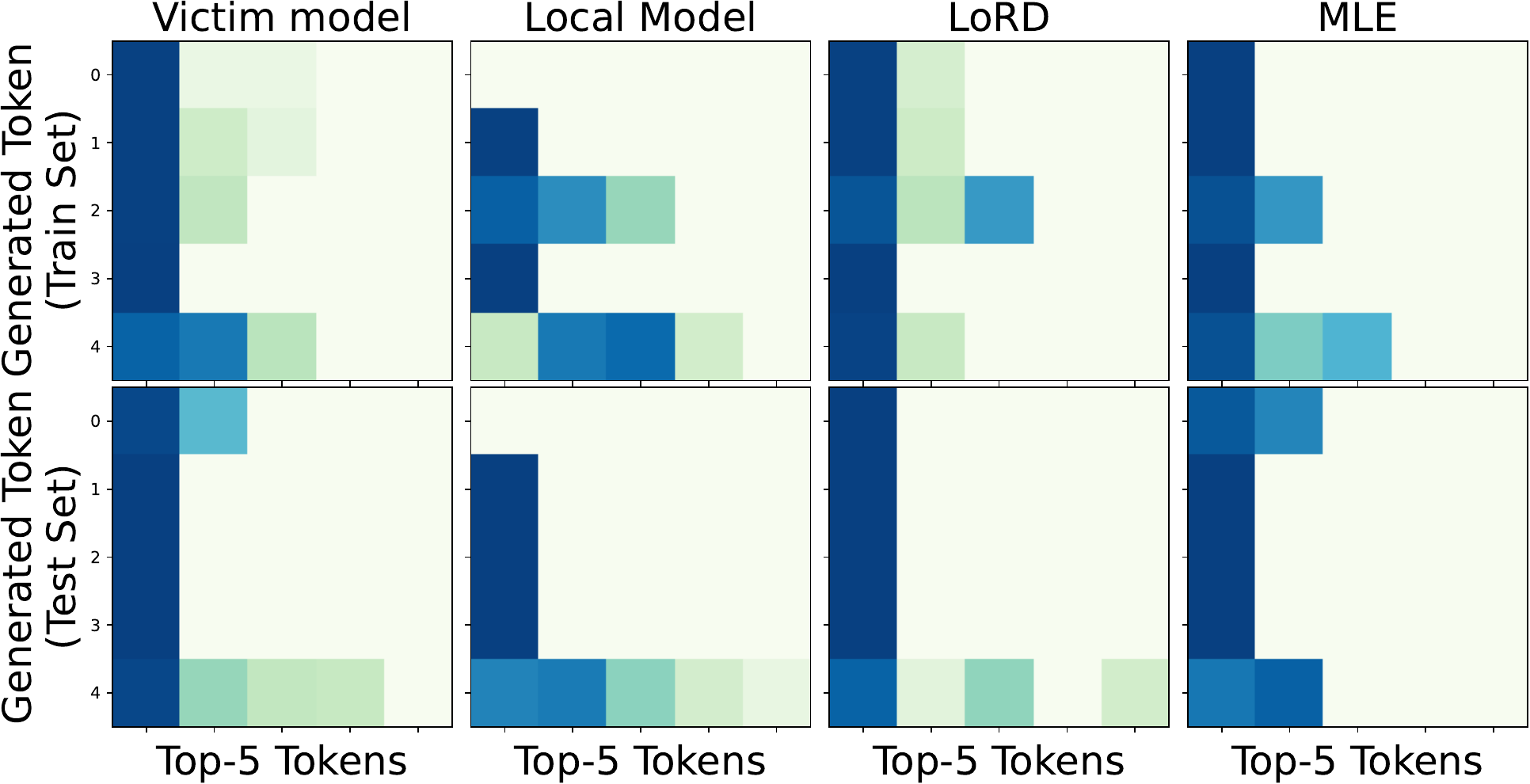}
\caption{Token generation distributions of four models, namely the victim model, the (initial) local model, and the local model learned through LoRD and MLE, respectively. We visualize their logarithmic probability on examples sampled from the train set and test set, where a deeper color indicates a higher probability.
}
\label{fig:distviz}
\end{figure}

We also investigate the \emph{probability distributions} in the generation
procedure among different extraction methods. Specifically, we
visualize these distributions for four models, the victim model
(GPT-3.5-turbo), the initial local model (llama3-8B), and the
learned local models with MLE and LoRD. As plotted in Figure
\ref{fig:distviz}, each row in the subfigures refers to the
distribution when generating the $i$-th token, with each column
element indicating the \emph{probability} predicted for the
corresponding token index.
We limit the visualization to no more than five token probabilities as currently only GPT-3.5-turbo provides the token prediction probabilities during generation, with a maximum of 5 candidate tokens ~\citep{openai-doc}.

From Figure \ref{fig:distviz}, we can see that both MLE and LoRD successfully redistribute the generation of the initial local model into a distribution similar to the victim model's, where probabilities, especially Top-1 tokens, have been well inherited in the extraction. This phenomenon supports our analysis in Proposition \ref{th:eqc}. However, distributions of MLE extracted models are consistently sharper than LoRD's, which aligns with our analysis in Section \ref{sec:q-eff}, where we claim that MLE leads local models to overfit to the preferred sentences (i.e., Top-1 tokens), namely \emph{PO}, and thus to disrupt the original distributions, leveraging unusual low probabilities for other token indexes. The reason why LoRD can be resistant watermarks, i.e., tokens in Top-1, can also be derived from this discovery.

\begin{table*}[t]
\centering
\resizebox{0.75\linewidth}{!}{%
\begin{tabular}{l|rrrrrrrr}
\Xhline{1.5pt}
\multicolumn{1}{c|}{\multirow{2}{*}{Method}} & \multicolumn{4}{c}{BLEU} & \multicolumn{3}{c}{BERTScore} & \multicolumn{1}{c}{Rouge-L} \\
\multicolumn{1}{c|}{} & \multicolumn{1}{c}{1} & \multicolumn{1}{c}{2} & \multicolumn{1}{c}{3} & \multicolumn{1}{c}{4} & \multicolumn{1}{c}{Precision} & \multicolumn{1}{c}{Recall} & \multicolumn{1}{c}{F1} & \multicolumn{1}{c}{F1} \\ \hline
LoRD~~\,(Q=16)~~~(T=0.8)& 54.40 & 42.18 & 33.56 & 27.06 & 89.89 & 94.06 & 91.44 & 56.09 \\ \hline
SimPO (Q=16)~~~(T=1.0) & 44.80 & 34.80 & 27.94 & 22.83 & 89.79 & 93.50 & 91.57 & 48.39 \\
SimPO (Q=16)~~~(T=1.3) & 44.19 & 33.45 & 26.31 & 21.18 & 88.49 & 92.65 & 90.47 & 47.09 \\
SimPO (Q=16)~~~(T=0.8) & 42.99 & 31.81 & 24.85 & 19.82 & 90.37 & 88.32 & 92.64 & 44.04 \\ \hline
SimPO (Q=256) (T=1.3) & 3.09 & 0.13 & 0.00 & 0.00 & 68.04 & 81.54 & 74.17 & 11.22 \\
SimPO (Q=256) (T=0.8) & 20.99 & 10.75 & 7.01 & 5.04 & 85.56 & 87.52 & 86.50 & 21.08 \\ \Xhline{1.35pt}
\end{tabular}%
}
\caption{Comparison between LoRD and The Direct Prompting with
  SimPO. $T$ denotes the temperature of local model's sampling, and $Q$ denotes the query times.}
\label{tab:vs-direct}
\end{table*}

\begin{table}
\centering
\resizebox{0.48\textwidth}{!}{%
\begin{tabular}{lrrrrr}
\Xhline{1.5pt}
\multicolumn{1}{c|}{\multirow{2}{*}{Models\textbackslash{}Metrics}} &
  \multicolumn{1}{c}{\multirow{2}{*}{Entropy}} &
  \multicolumn{2}{c}{ To Victim Model} &
  \multicolumn{2}{c}{To Initial Local Model} \\ \cline{3-6}
\multicolumn{1}{c|}{} &
  \multicolumn{1}{c}{} &
  \multicolumn{1}{c}{$\mathbb{D}_{\text{KL}}\downarrow$} &
  \multicolumn{1}{c}{Spear. Corr.$\uparrow$} &
  \multicolumn{1}{c}{$\mathbb{D}_{\text{KL}}$} &
  \multicolumn{1}{c}{Spear. Corr.} \\ \hline
\multicolumn{6}{c}{\emph{On training dataset}}             \\ \hline
\multicolumn{1}{l|}{Initial Local Model} & 0.395 & 0.503 & 0.620 & -     & -     \\
\multicolumn{1}{l|}{+ LoRD}              & 0.209 & 0.051 & \textbf{0.880} & 0.169 & 0.680 \\
\multicolumn{1}{l|}{+ MLE}               & 0.271 & \textbf{0.029} & 0.780 & 0.051 & 0.540 \\ \hline
\multicolumn{6}{c}{\emph{On the test dataset}}          \\ \hline
\multicolumn{1}{l|}{Initial Local Model} & 0.269 & 0.471 & 0.680 & -     & -     \\
\multicolumn{1}{l|}{+ LoRD}              & 0.122 & 0.033 & \textbf{0.640} & 0.046 & 0.720 \\
\multicolumn{1}{l|}{+MLE}                & 0.275 & \textbf{0.032} & 0.274 & 0.001 & 0.740 \\ \hline
\Xhline{1pt}
\end{tabular}%
}
  \caption{Quantization analysis on distributions.
    A low KL divergence or a high Spearman correlation indicates a high
    similarity.}
\label{tab:viz}
\end{table}

To compare MLE and LoRD accurately, we quantize the \emph{entropy} of these distributions, and compute the \emph{KL divergence} ($\mathbb{D}_{\text{KL}}$), and the \emph{Spearman
  Correlation (Spear. Corr.)} with respect to the victim and initial local
model. As shown in Table \ref{tab:viz}, while the MLE extracted model
exhibits a lower KL divergence (i.e., high distribution similarity) with
the victim model than LoRD's on the training dataset,
its KL divergence becomes comparable to LoRD's on the test
set. Meanwhile, its Spearman correlation significantly
decreases from 0.78 to 0.27, which shows that MLE cannot effectively
imitate victim model's prediction behaviors when encountering data
beyond the training dataset.

\subsection{Ablation Study}

We conduct an ablation study to assess the impact of our proposed loss
functions shown in Section \ref{sec:lord}. Specifically, we
adopt the same experimental settings described in Section
\ref{sec:set} and compare LoRD against the following variations on the
WMT16 (de-en) dataset:

\noindent
$\bullet$ \textbf{w.o.} $\sigma(\cdot)$: Removing the sigmoid function in Equation \ref{eq:black};

\noindent
$\bullet$ \textbf{Rep.} $\mathbf{y}^{-}$ \textbf{w.} $\mathbf{y}^{+}$: Replacing $\mathbf{y}^{-}_{t-1}$ with $\mathbf{y}^{+}_{t-1}$ defined in Equation \ref{eq:reg};

\noindent
$\bullet$ \textbf{w.o.} $\mathcal{L}_{reg}$: Eliminating the regularization term.

The ablation results are presented in Table \ref{tab:ablation}. Our
findings indicate that the sigmoid function used for normalization is
not essential for the effectiveness of our extraction
strategy. However, the regularization term proves to be crucial for
ensuring the model's convergence, which is consistent with our
theoretical analysis.

\subsection{LoRD versus Direct Prompting}

We notice that the victim model can serve as a feedback signal to explicitly determine $\mathbf{y}^{+}_{t-1}$ and $\mathbf{y}^{-}_{t-1}$, thereby enabling a reinforcement learning (RL) approach based on direct prompting. This idea aligns with prior work on reinforcement learning with AI feedback (RLAIF), as discussed in Appendix \ref{sec:related-rlaif}.

In this section, we present an empirical comparison between LoRD and
direct prompting and argue that direct prompting is less suitable for
MEAs than our LoRD.

\begin{table}[t]
\centering
\resizebox{\linewidth}{!}{%
\begin{tabular}{l|rrrrr}
  \Xhline{1.5pt}
\multicolumn{1}{c|}{\multirow{2}{*}{Method}} & \multicolumn{1}{c}{\multirow{2}{*}{BLEU-4}} & \multicolumn{3}{c}{BERTScore} & \multicolumn{1}{c}{Rouge-L} \\
\multicolumn{1}{c|}{} & \multicolumn{1}{c}{} & \multicolumn{1}{c}{Precision} & \multicolumn{1}{c}{Recall} & \multicolumn{1}{c}{F1} & \multicolumn{1}{c}{F1} \\
  \hline
LoRD & 27.06 & 89.89 & 94.06 & 91.44 & 56.09 \\
w.o. $\sigma(\cdot)$ & 23.77 & 89.25 & 93.73 & 91.38 & 50.39 \\
Rep. $\mathbf{y}^-_{t-1}$ w. $\mathbf{y}^+_{t-1}$ & 25.87 & 87.41 & 93.28 & 90.19 & 54.12 \\
 w.o. $\mathcal{L}_{reg}$ & NC & NC & NC & NC & NC \\
  \Xhline{1.35pt}
\end{tabular}%
}
\caption{Ablation Study for LoRD. NC denotes that the model does not
  converged during training.}
\label{tab:ablation}
\end{table}

\noindent
\textbf{Empirical Comparison.}
We design a prompt to obtain feedback from the victim model as
``\texttt{For a translation task involving the conversion of the given
  `Text` into English, the user will provide two translation versions
  labeled `A` and `B`. Your task is to return the *letter
  corresponding to the better translation* without including any
  additional output.}''.
For direct prompting, we allow the victim model to determine the
positive and negative responses generated by the local model. These
responses are then used to fine-tune the local model using a
DPO-inspired loss function. Specifically, we employ SimPO~\cite{simpo}
as the loss function. To ensure a fair comparison, we maintain the
same hyperparameter settings as in previous experiments.

As shown in Table \ref{tab:vs-direct}, we conducted experiments with
various sampling temperatures for the direct
prompting. However, the performance of the direct prompting still
underperforms LoRD. This limitation may stem from the local model's
lack of guidance from correct answers. When the local model generates
two suboptimal responses, a direct prompting-based method is compelled
to select the "winner" of two inadequate response rather than an
optimal response, which we believe is the crux of the issue.

\begin{table*}[t]
\centering
\resizebox{0.98\textwidth}{!}{%
\begin{tabular}{r|rrrrrrrrrr}
\Xhline{2.5pt}
             & \multicolumn{4}{c}{BLEU} & \multicolumn{3}{c}{BERTScore} & \multicolumn{3}{c}{Rouge-L} \\\cline{2-11}
            & 1 & 2 & 3 & 4 & Pre. & Rec. & F1.  & Pre. & Rec. & F1. \\\hline
\multicolumn{11}{c}{\emph{Text to SQL: WikiSQL~\citep{wikisql} with 64 query samples}}                                                                             \\\hline
Victim Model  &\greyc 54.1 & \greyc 41.4 & \greyc 32.1  & \greyc 24.4 & \greyc 86.9 & \greyc 93.5 & \greyc 90.1 & \greyc 58.9 & \greyc 62.1 & \greyc 59.7 \\
Local Model &\red $20.2\pm 0.2$ &\red  $14.5\pm 0.2$& \red $10.9\pm 0.1$&\red $8.1\pm 0.1$&\gres $82.5\pm 0.0$&$\grem 92.4\pm 0.1$&\gres $87.1\pm 0.0$&\red $22.6\pm 0.3$ & \grexl $66.4\pm 0.4$& \red $33.2\pm 0.3$ \\
+MLE         & \grem $54.0\pm 1.6$ & \gres $37.5\pm 2.1$ & \gres $26.4\pm 2.0$ & \gret $18.8\pm 1.8$ &\gret $83.1\pm 0.2$ &\grem $92.9\pm 0.2$& \grem $87.7\pm 0.2$& \gres  $56.2\pm 1.5$& \gret $56.1\pm 0.9$   &\gret $55.8\pm 1.2$  \\
+LoRD         &\grel  $55.1\pm 2.3$ &\grem  $39.0\pm 3.6$ &\grem  $28.0\pm 4.0$ &\gres  $20.4\pm 3.9$ &\gres $83.4\pm 0.4$ &\grem $92.9\pm 0.3$&\grem  $87.9\pm 0.4$&\gres   $57.7\pm 2.2$& \gret $56.3\pm 2.0$   &\gres $56.7\pm 2.1$  \\\hline
\multicolumn{11}{c}{\emph{Text to SQL: Spider~\citep{wikisql} with 64 query samples}}                                                                             \\\hline
Victim Model & \greyc 9.4 & \greyc 3.9 & \greyc 2.1 & \greyc 1.1 & \greyc 77.7& \greyc 84.1 &\greyc 80.6 & \greyc 17.1& \greyc 36.3& \greyc 21.8\\
Local Model & \gres $6.4\pm 0.2$ & \gres  $2.1\pm 0.1$& \grem  $0.9\pm 0.1$& \grel  $0.5\pm 0.0$& \grexl  $80.0\pm 0.1$&\grem  $82.6\pm 0.1$& \grel  $81.2\pm 0.1$& \gret  $10.0\pm 0.3$&\red  $21.5\pm 0.6$ & \gret  $12.7\pm 0.4$ \\
+MLE         & \gres   $6.2\pm 0.9$ & \gres  $1.3\pm 0.5$ & \grem  $0.6\pm 0.3$ & \grel  $0.2\pm 0.2$ & \grem $76.4\pm 0.7$ & \gres $81.8\pm 0.4$& \grem  $78.9\pm 0.6$& \gres   $12.7\pm 1.6$& \red  $18.3\pm 1.6$   & \gret $14.3\pm 1.6$  \\
+LoRD         & \grel   $9.1\pm 0.9$ & \grem  $2.8\pm 0.5$ & \grel  $1.3\pm 0.4$ & \grel  $0.6\pm 0.2$ & \grel $77.7\pm 0.4$ & \grel $83.1\pm 0.5$& \grel  $80.2\pm 0.3$& \grel   $16.9\pm 0.1$& \gret  $24.1\pm 0.2$   & \gres $18.8\pm 0.1$  \\\hline
\multicolumn{11}{c}{\emph{Data to Text: E2E NLG~\citep{e2enlg} with 64 query samples}}                                                                             \\\hline
Victim Model & \greyc 51.8& \greyc 27.0& \greyc 26.8& \greyc 19.1& \greyc 93.9& \greyc 94.6& \greyc 94.2& \greyc 49.6 & \greyc 54.6& \greyc 51.4\\
Local Model  & \red  $31.1\pm 0.1$& \gret  $20.1\pm 0.2$& \red  $13.5\pm 0.2$& \red  $8.9\pm 0.3$& \gret  $86.1\pm 0.1$& \grem  $92.4\pm 0.1$& \gres  $89.1\pm 0.1$& \red   $29.0\pm 0.3$& \gret  $49.4\pm 0.4$& \gret  $35.9\pm 0.3$\\
+MLE         & \grexl  $53.0\pm 0.9$ & \grel   $38.0\pm 0.6$& \grel $27.5\pm 0.5$ & \grel $19.9\pm 0.4$ & \gres $89.1\pm 0.0$ & \grel $94.5\pm 0.0$ & \gres $91.8\pm 0.0$ & \grem  $48.3\pm 0.5$& \grel $54.2\pm 1.4$ & \grem $50.4\pm 0.9$ \\
+LoRD         & \grexl  $53.1\pm 1.1$ & \grel   $38.2\pm 0.9$& \grel $27.8\pm 0.7$ & \grexl $20.2\pm 0.5$ & \gres $89.1\pm 0.1$ & \grel $94.5\pm 0.1$ & \gres $91.7\pm 0.1$ & \grem  $48.3\pm 0.7$& \grem $53.5\pm 1.4$ & \grem $50.2\pm 0.9$ \\\hline
\multicolumn{11}{c}{\emph{Data to Text: CommonGen~\citep{commongen} with 64 query samples}}                                                                             \\\hline
Victim Model & \greyc  33.3& \greyc 18.5 & \greyc 11.1& \greyc 6.9& \greyc 91.3& \greyc 92.1& \greyc 91.7& \greyc 33.6 &\greyc 40.7 & \greyc 36.1\\
Local Model  &\red  $12.2\pm 0.0$ &\red  $6.5\pm 0.1$ & \gret  $3.8\pm 0.0$ & \gret  $2.3\pm 0.0$ & \gret  $83.0\pm 0.0$ & \gres  $89.7\pm 0.0$ & \gres  $86.2\pm 0.0$ & \red  $14.6\pm 0.1$ & \grexl  $46.2\pm 0.2$ & \red  $21.6\pm 0.0$ \\
+MLE         & \grem $32.4\pm 2.0$ & \grel $18.3\pm 1.3$ & \grel $10.9\pm 1.0$ & \grel $6.6\pm 0.7$ & \gret $84.2\pm 0.1$ & \grel $91.7\pm 0.0$ & \gres $87.8\pm 0.0$ & \grem $31.7\pm 2.4$ & \grel $41.1\pm 0.4$ & \grem $35.1\pm 1.6$ \\
+LoRD         & \grem $32.1\pm 1.3$ & \grel $18.0\pm 0.9$ & \grel $10.7\pm 0.5$ & \grel $6.4\pm 0.3$ & \gret $84.1\pm 0.0$ & \grel $91.6\pm 0.1$ & \gres $87.7\pm 0.0$ & \grem $31.4\pm 1.1$ & \grel $40.3\pm 0.9$ & \grem $34.6\pm 0.9$ \\\hline
\multicolumn{11}{c}{\emph{Summarization: TLDR~\citep{tldr} with 64 query samples}}                                                                             \\\hline
Victim Model & \greyc 11.9& \greyc 5.0 & \greyc 2.6 & \greyc 1.5& \greyc 85.9& \greyc 88.4& \greyc 87.1& \greyc 13.4&\greyc 30.9& \greyc 18.4\\
Local Model  & \red  $6.9\pm 0.0$& \grem  $3.2\pm 0.1$& \grel  $1.7\pm 0.0$& \grel  $1.0\pm 0.0$& \gres  $81.0\pm 0.1$& \grem  $87.6\pm 0.0$& \gres  $84.1\pm 0.0$& \gres  $10.5\pm 0.1$& \grel  $41.1\pm 0.1$& \red  $16.4\pm 0.1$\\
+MLE         & \grel  $10.6\pm 0.5$& \grel $4.8\pm 0.2$ & \grel $2.6\pm 0.1$ & \grel $1.6\pm 1.1$ & \grem $83.6\pm 0.7$ & \grel $88.4\pm 0.2$ & \grem $85.9\pm 0.5$ & \grel $14.3\pm 0.5$ & \grexl $32.7\pm 1.1$& \grel  $18.9\pm 0.4$\\
+LoRD         & \grem  $10.2\pm 0.3$& \grel $4.5\pm 0.1$ & \grel $2.4\pm 0.1$ & \grel $1.4\pm 0.0$ & \grem $84.1\pm 0.1$ & \grel $88.3\pm 0.1$ & \grel $86.2\pm 0.1$ & \grel $12.8\pm 0.3$ & \grexl $33.2\pm 0.9$& \grel  $18.0\pm 0.2$\\\hline
\multicolumn{11}{c}{\emph{Summarization: CNN Daily Mail~\citep{cnn} with 64 query samples}}                                                                             \\\hline
Victim Model & \greyc 20.4 & \greyc 10.8& \greyc 6.4& \greyc 4.1&\greyc 86.4 & \greyc 87.8& \greyc 87.1& \greyc 22.4& \greyc 40.8& \greyc 28.2\\
Local Model  & \red  $4.9\pm 0.0$& \gret  $3.6\pm 0.0$& \gres  $2.7\pm 0.0$& \grem  $2.1\pm 0.0$& \gret $80.5\pm 0.0$& \grel  $88.3\pm 0.0$& \gres   $84.2\pm 0.0$& \red  $10.9\pm 0.0$& \grexl  $79.1\pm 0.1$& \gret  $18.8\pm 0.0$\\
+MLE         & \red  $5.1\pm 0.5$& \gret  $3.7\pm 0.0$& \gres  $2.8\pm 0.0$& \grem  $2.2\pm 0.0$& \gret  $80.6\pm 0.0$& \grexl  $88.3\pm 0.0$& \gres  $84.3\pm 0.0$& \red  $11.3\pm 0.1$& \grexl  $78.6\pm 0.1$  & \red  $19.3\pm 0.1$\\
+LoRD         & \red  $5.3\pm 0.0$& \gret  $3.9\pm 0.0$& \gres  $2.9\pm 0.0$& \grem  $2.3\pm 0.0$& \gret  $80.6\pm 0.0$& \grexl  $88.4\pm 0.0$& \gres  $84.3\pm 0.0$& \red  $11.3\pm 0.1$& \grexl  $78.6\pm 0.2$  & \red   $19.1\pm 0.1$\\\hline
\multicolumn{11}{c}{\emph{Summarization: Samsum~\citep{samsum} with 64 query samples}}                                                                             \\\hline
Victim Model & \greyc 20.7& \greyc 11.4& \greyc 6.9& \greyc 4.4& \greyc 88.1& \greyc 91.7& \greyc 89.8& \greyc 24.2& \greyc 50.5& \greyc 31.6\\
Local Model  & \red  $8.9\pm 0.2$& \gret  $5.2\pm 0.1$& \gres   $3.3\pm 0.1$& \grem  $2.1\pm 0.1$& \gret  $80.9\pm 0.2$& \grem  $90.1\pm 0.1$& \gres  $85.2\pm 0.2$& \gret  $17.0\pm 0.3$& \grexl  $61.8\pm 0.5$& \gret  $25.5\pm 0.4$\\
+MLE         & \gres  $16.9\pm 1.1$& \gres  $9.4\pm 0.7$& \grem  $5.8\pm 0.4$& \grel  $3.7\pm 0.3$& \gret  $83.9\pm 0.9$& \grel  $90.9\pm 0.6$& \grem  $87.3\pm 0.8$& \grel  $25.2\pm 0.8$& \grel  $49.8\pm 2.5$  & \grel   $31.0\pm 1.7$\\
+LoRD        & \grem  $18.4\pm 0.7$& \grem  $10.1\pm 0.3$& \grel  $6.0\pm 0.2$& \grel  $3.7\pm 0.1$& \gres  $84.9\pm 0.1$& \grel  $91.5\pm 0.1$& \grem  $88.1\pm 0.1$& \grel  $23.2\pm 0.8$& \grel  $49.7\pm 1.5$  & \grem   $30.2\pm 0.6$\\
\Xhline{1.5pt}
\end{tabular}%
}
\caption{MEA comparison on three tasks, including structured text
  generation, data to text, and summarization.
  We use GPT-3.5-turbo as the victim model, and Llama3-8B~\citep{llama3} as the
local initial model.
The \emph{intensity} of the red or blue color corresponds to the
degree of underperformance or outperformance relative to the victim
model.
}
\label{tab:domain-res}
\end{table*}

RLHF tackles this challenge by incorporating a regularization term
with the initial model, LoRD addresses it through our $\mathcal{L}_{reg}$,
and DPO resolves it by employing the training corpus of the reward
model. Unfortunately, a direct prompt-based method overlooks this
point. To further investigate this problem, we increased the query
number to 256, which resulted in the local model failing to converge
and exhibiting poor performance.

Besides, we also observed a bias in the victim model's selection
between the first and second sentences. In a series of 256 queries,
the model successfully provided an answer (either A or B) 255
times. However, it chose the first sentence only 84 times, which is a
mere 32.94\%, significantly deviating from the expected 50\%. Given
that the generated sentences are randomly sampled from the local model
without any significant correlation to their order, we deduce that
relying on the victim model to directly generate feedback might be, at
least, an unreliable approach. It may necessitate additional
considerations for the design of the prompt and the capabilities of
the victim model to ensure the robustness of these algorithms.

\begin{figure*}[t]
\includegraphics[width=0.99\linewidth]{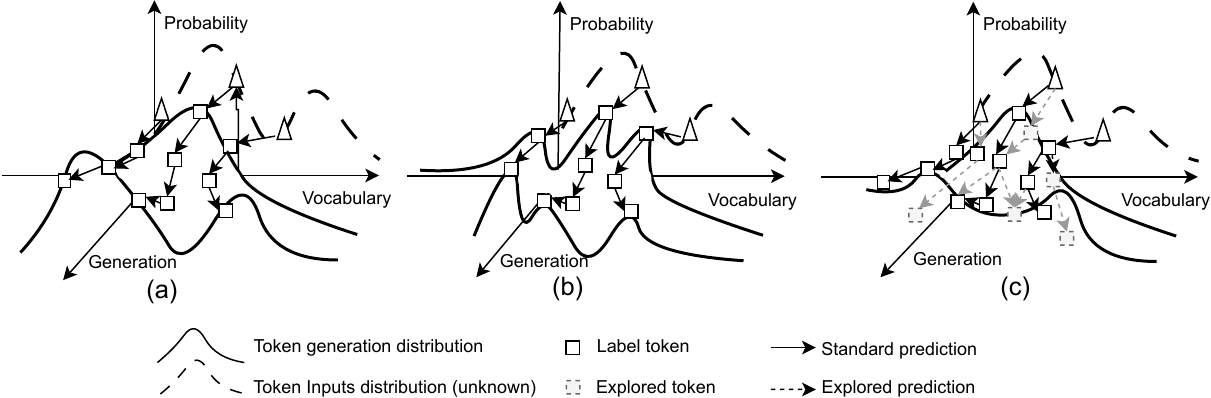}
\caption{Comparison of learned joint \emph{prediction distributions} among the victim model (a), local models are learned with MLE (b) and LoRD (c). Simply obtaining the tokens from the victim model (solid black squares), MLE may only memorize specific responses and build a complicated decision surface, resulting in \emph{preference overfitting}. In contrast, LoRD further explores the candidate generation paths (dashed arrows and squares) under the guidance of the victim's generation, which is expected to better approximate the victim model in terms of generalization ability, especially under a limited query budget.
}
\label{fig:intro2}
\end{figure*}

\noindent
\textbf{Discussion on the Feasibility.}
In addition to the empirical comparison, we provide a discussion
supporting the proposition that direct prompting is unsuitable for
model extraction attacks for the following reasons:
\begin{itemize}
\item A direct feedback query will \emph{expose the intention of the adversary};
\item Unlike the current design of LoRD, direct feedback is contingent
  upon the local model's responses, which is
  query-\textbf{in}efficient. Specifically, for a given query sample, the
  algorithm would need to repeatedly query the victim model to
  distinguish between $\mathbf{y}^{+}_{t-1}$ and
  $\mathbf{y}^{-}_{t-1}$ across different learning periods. On the
  contrary, LoRD necessitates only a single query per sample to
  discriminate different $(\mathbf{y}^{+}_{t-1},\mathbf{y}^{-}_{t-1})$
  pairs;
\item The threat model will change if employing a direct prompting. As we
  know, both LoRD and MLE are currently trained under the same
  conditions, i.e. $(\mathbf{x},\mathbf{y}_{vic})$ paires. The fairness would be questioned when we compare methods under disparate query settings.
\end{itemize}


%% file: main.bbl
\begin{thebibliography}{71}
\providecommand{\natexlab}[1]{#1}

\bibitem[{Achiam et~al.(2024)Achiam, Adler, Agarwal, Ahmad, Akkaya, Aleman,
  Almeida, Altenschmidt, Altman, Anadkat, Avila, Babuschkin, Balaji, Balcom,
  Baltescu, Bao, Bavarian, Belgum, Bello, Berdine, Bernadett-Shapiro, Berner,
  Bogdonoff, Boiko, Boyd, and et~al.}]{gpt4}
Josh Achiam, Steven Adler, Sandhini Agarwal, Lama Ahmad, Ilge Akkaya,
  Florencia~Leoni Aleman, Diogo Almeida, Janko Altenschmidt, Sam Altman,
  Shyamal Anadkat, Red Avila, Igor Babuschkin, Suchir Balaji, Valerie Balcom,
  Paul Baltescu, Haiming Bao, Mohammad Bavarian, Jeff Belgum, Irwan Bello, Jake
  Berdine, Gabriel Bernadett-Shapiro, Christopher Berner, Lenny Bogdonoff, Oleg
  Boiko, Madelaine Boyd, and Anna-Luisa~Brakman et~al. 2024.
\newblock \href {https://arxiv.org/abs/2303.08774} {Gpt-4 technical report}.
\newblock \emph{Preprint}, arXiv:2303.08774.

\bibitem[{Anil et~al.(2024)Anil, Borgeaud, Alayrac, Yu, Soricut, Schalkwyk, and
  et~al.}]{gemini}
Rohan Anil, Sebastian Borgeaud, Jean-Baptiste Alayrac, Jiahui Yu, Radu Soricut,
  Johan Schalkwyk, and Andrew M.~Dai et~al. 2024.
\newblock \href {https://arxiv.org/abs/2312.11805} {Gemini: A family of highly
  capable multimodal models}.
\newblock \emph{Preprint}, arXiv:2312.11805.

\bibitem[{Anthropic(2024)}]{claude}
Anthropic. 2024.
\newblock \href {https://api.semanticscholar.org/CorpusID:268232499} {The
  claude 3 model family: Opus, sonnet, haiku}.

\bibitem[{Bai et~al.(2022{\natexlab{a}})Bai, Jones, Ndousse, Askell, Chen,
  DasSarma, Drain, Fort, Ganguli, Henighan, Joseph, Kadavath, Kernion, Conerly,
  El-Showk, and et~al.}]{rlhf1}
Yuntao Bai, Andy Jones, Kamal Ndousse, Amanda Askell, Anna Chen, Nova DasSarma,
  Dawn Drain, Stanislav Fort, Deep Ganguli, Tom Henighan, Nicholas Joseph,
  Saurav Kadavath, Jackson Kernion, Tom Conerly, Sheer El-Showk, and
  Nelson~Elhage et~al. 2022{\natexlab{a}}.
\newblock \href {https://arxiv.org/abs/2204.05862} {Training a helpful and
  harmless assistant with reinforcement learning from human feedback}.
\newblock \emph{Preprint}, arXiv:2204.05862.

\bibitem[{Bai et~al.(2022{\natexlab{b}})Bai, Kadavath, Kundu, Askell, Kernion,
  Jones, Chen, Goldie, Mirhoseini, McKinnon, and et~al.}]{rlhf2}
Yuntao Bai, Saurav Kadavath, Sandipan Kundu, Amanda Askell, Jackson Kernion,
  Andy Jones, Anna Chen, Anna Goldie, Azalia Mirhoseini, Cameron McKinnon, and
  Carol~Chen et~al. 2022{\natexlab{b}}.
\newblock \href {https://doi.org/10.48550/ARXIV.2212.08073} {Constitutional
  {AI:} harmlessness from {AI} feedback}.
\newblock \emph{CoRR}, abs/2212.08073.

\bibitem[{Bailey et~al.(2012)Bailey, Dittrich, Kenneally, and Maughan}]{menlo}
Michael Bailey, David Dittrich, Erin Kenneally, and Doug Maughan. 2012.
\newblock \href {https://doi.org/10.1109/MSP.2012.52} {The menlo report}.
\newblock \emph{IEEE Security and Privacy}, 10(2):71–75.

\bibitem[{Bengio et~al.(2000)Bengio, Ducharme, and Vincent}]{bengio2000neural}
Yoshua Bengio, R{\'e}jean Ducharme, and Pascal Vincent. 2000.
\newblock A neural probabilistic language model.
\newblock \emph{Advances in neural information processing systems}, 13.

\bibitem[{Bisk et~al.(2020)Bisk, Zellers, Bras, Gao, and Choi}]{piqa}
Yonatan Bisk, Rowan Zellers, Ronan~Le Bras, Jianfeng Gao, and Yejin Choi. 2020.
\newblock Piqa: Reasoning about physical commonsense in natural language.
\newblock In \emph{Thirty-Fourth AAAI Conference on Artificial Intelligence}.

\bibitem[{Bojar et~al.(2016)Bojar, Chatterjee, Federmann, Graham, Haddow, Huck,
  Jimeno~Yepes, Koehn, Logacheva, Monz, Negri, Neveol, Neves, Popel, Post,
  Rubino, Scarton, Specia, Turchi, Verspoor, and Zampieri}]{wmt16}
Ond~{r}ej Bojar, Rajen Chatterjee, Christian Federmann, Yvette Graham, Barry
  Haddow, Matthias Huck, Antonio Jimeno~Yepes, Philipp Koehn, Varvara
  Logacheva, Christof Monz, Matteo Negri, Aurelie Neveol, Mariana Neves, Martin
  Popel, Matt Post, Raphael Rubino, Carolina Scarton, Lucia Specia, Marco
  Turchi, Karin Verspoor, and Marcos Zampieri. 2016.
\newblock \href {http://www.aclweb.org/anthology/W/W16/W16-2301} {Findings of
  the 2016 conference on machine translation}.
\newblock In \emph{Proceedings of the First Conference on Machine Translation},
  pages 131--198, Berlin, Germany. Association for Computational Linguistics.

\bibitem[{Cong et~al.(2022)Cong, He, and Zhang}]{wm1}
Tianshuo Cong, Xinlei He, and Yang Zhang. 2022.
\newblock \href {https://doi.org/10.1145/3548606.3559355} {Sslguard: {A}
  watermarking scheme for self-supervised learning pre-trained encoders}.
\newblock In \emph{Proceedings of the 2022 {ACM} {SIGSAC} Conference on
  Computer and Communications Security, {CCS} 2022, Los Angeles, CA, USA,
  November 7-11, 2022}, pages 579--593. {ACM}.

\bibitem[{Devlin et~al.(2019)Devlin, Chang, Lee, and Toutanova}]{bert}
Jacob Devlin, Ming-Wei Chang, Kenton Lee, and Kristina Toutanova. 2019.
\newblock \href {https://doi.org/10.18653/v1/N19-1423} {{BERT}: Pre-training of
  deep bidirectional transformers for language understanding}.
\newblock In \emph{Proceedings of the 2019 Conference of the North {A}merican
  Chapter of the Association for Computational Linguistics: Human Language
  Technologies, Volume 1 (Long and Short Papers)}, pages 4171--4186,
  Minneapolis, Minnesota. Association for Computational Linguistics.

\bibitem[{Du{\v{s}}ek et~al.(2020)Du{\v{s}}ek, Novikova, and Rieser}]{e2enlg}
Ond\v{r}ej Du{\v{s}}ek, Jekaterina Novikova, and Verena Rieser. 2020.
\newblock \href {https://doi.org/10.1016/j.csl.2019.06.009} {Evaluating the
  {{State}}-of-the-{{Art}} of {{End}}-to-{{End Natural Language Generation}}:
  {{The E2E NLG Challenge}}}.
\newblock \emph{Computer Speech \& Language}, 59:123--156.

\bibitem[{Glaese et~al.(2022)Glaese, McAleese, Trebacz, Aslanides, Firoiu,
  Ewalds, Rauh, Weidinger, Chadwick, Thacker, Campbell{-}Gillingham, Uesato,
  Huang, Comanescu, and et~al.}]{sparrow}
Amelia Glaese, Nat McAleese, Maja Trebacz, John Aslanides, Vlad Firoiu, Timo
  Ewalds, Maribeth Rauh, Laura Weidinger, Martin Chadwick, Phoebe Thacker, Lucy
  Campbell{-}Gillingham, Jonathan Uesato, Po{-}Sen Huang, Ramona Comanescu, and
  Fan~Yang et~al. 2022.
\newblock Improving alignment of dialogue agents via targeted human judgements.
\newblock \emph{CoRR}, abs/2209.14375.

\bibitem[{Gliwa et~al.(2019)Gliwa, Mochol, Biesek, and Wawer}]{samsum}
Bogdan Gliwa, Iwona Mochol, Maciej Biesek, and Aleksander Wawer. 2019.
\newblock \href {https://doi.org/10.18653/v1/D19-5409} {{SAMS}um corpus: A
  human-annotated dialogue dataset for abstractive summarization}.
\newblock In \emph{Proceedings of the 2nd Workshop on New Frontiers in
  Summarization}, pages 70--79, Hong Kong, China. Association for Computational
  Linguistics.

\bibitem[{Go et~al.(2023)Go, Korbak, Kruszewski, Rozen, Ryu, and
  Dymetman}]{rl-deduct3}
Dongyoung Go, Tomasz Korbak, Germ{\'{a}}n Kruszewski, Jos Rozen, Nahyeon Ryu,
  and Marc Dymetman. 2023.
\newblock \href {https://proceedings.mlr.press/v202/go23a.html} {Aligning
  language models with preferences through f-divergence minimization}.
\newblock In \emph{International Conference on Machine Learning, {ICML} 2023,
  23-29 July 2023, Honolulu, Hawaii, {USA}}, volume 202 of \emph{Proceedings of
  Machine Learning Research}, pages 11546--11583. {PMLR}.

\bibitem[{Grattafiori et~al.(2024)Grattafiori, Dubey, and et~al.}]{llama3}
Aaron Grattafiori, Abhimanyu Dubey, and Abhinav~Jauhri et~al. 2024.
\newblock \href {https://arxiv.org/abs/2407.21783} {The llama 3 herd of
  models}.
\newblock \emph{Preprint}, arXiv:2407.21783.

\bibitem[{Gu et~al.(2022)Gu, Huang, Zheng, Chang, and Hsieh}]{model-wm2}
Chenxi Gu, Chengsong Huang, Xiaoqing Zheng, Kai-Wei Chang, and Cho-Jui Hsieh.
  2022.
\newblock \href {https://arxiv.org/abs/2210.07543} {Watermarking pre-trained
  language models with backdooring}.
\newblock \emph{Preprint}, arXiv:2210.07543.

\bibitem[{He et~al.(2021)He, Xu, Lyu, Wu, and Wang}]{wm4}
Xuanli He, Qiongkai Xu, Lingjuan Lyu, Fangzhao Wu, and Chenguang Wang. 2021.
\newblock \href {https://arxiv.org/abs/2112.02701} {Protecting intellectual
  property of language generation apis with lexical watermark}.
\newblock \emph{Preprint}, arXiv:2112.02701.

\bibitem[{He et~al.(2022)He, Xu, Zeng, Lyu, Wu, Li, and Jia}]{wm2}
Xuanli He, Qiongkai Xu, Yi~Zeng, Lingjuan Lyu, Fangzhao Wu, Jiwei Li, and Ruoxi
  Jia. 2022.
\newblock \href
  {http://papers.nips.cc/paper\_files/paper/2022/hash/2433fec2144ccf5fea1c9c5ebdbc3924-Abstract-Conference.html}
  {{CATER:} intellectual property protection on text generation apis via
  conditional watermarks}.
\newblock In \emph{Advances in Neural Information Processing Systems 35: Annual
  Conference on Neural Information Processing Systems 2022, NeurIPS 2022, New
  Orleans, LA, USA, November 28 - December 9, 2022}.

\bibitem[{Heath(2023)}]{company-steal}
Alex Heath. 2023.
\newblock Bytedance is secretly using openai’s tech to build a competitor.
\newblock [Online].
\newblock
  \url{https://www.theverge.com/2023/12/15/24003151/bytedance-china-openai-microsoft-competitor-llm}.

\bibitem[{Hermann et~al.(2015)Hermann, Kociský, Grefenstette, Espeholt, Kay,
  Suleyman, and Blunsom}]{cnn}
Karl~Moritz Hermann, Tomás Kociský, Edward Grefenstette, Lasse Espeholt, Will
  Kay, Mustafa Suleyman, and Phil Blunsom. 2015.
\newblock \href
  {http://papers.nips.cc/paper/5945-teaching-machines-to-read-and-comprehend}
  {Teaching machines to read and comprehend}.
\newblock In \emph{NIPS}, pages 1693--1701.

\bibitem[{Hinton et~al.(2015)Hinton, Vinyals, and Dean}]{kd}
Geoffrey~E. Hinton, Oriol Vinyals, and Jeffrey Dean. 2015.
\newblock \href {https://arxiv.org/abs/1503.02531} {Distilling the knowledge in
  a neural network}.
\newblock \emph{CoRR}, abs/1503.02531.

\bibitem[{Jagielski et~al.(2020)Jagielski, Carlini, Berthelot, Kurakin, and
  Papernot}]{mea-tra-high}
Matthew Jagielski, Nicholas Carlini, David Berthelot, Alex Kurakin, and Nicolas
  Papernot. 2020.
\newblock High accuracy and high fidelity extraction of neural networks.
\newblock In \emph{29th USENIX Security Symposium (USENIX Security 20)}, pages
  1345--1362.

\bibitem[{Ji et~al.(2024)Ji, Hong, Zhang, Chen, Dai, Zheng, Qiu, Li, and
  Yang}]{saferlhf}
Jiaming Ji, Donghai Hong, Borong Zhang, Boyuan Chen, Josef Dai, Boren Zheng,
  Tianyi Qiu, Boxun Li, and Yaodong Yang. 2024.
\newblock Pku-saferlhf: Towards multi-level safety alignment for llms with
  human preference.
\newblock \emph{arXiv preprint arXiv:2406.15513}.

\bibitem[{Jia et~al.(2021)Jia, Choquette{-}Choo, Chandrasekaran, and
  Papernot}]{backdoor-mea-wm}
Hengrui Jia, Christopher~A. Choquette{-}Choo, Varun Chandrasekaran, and Nicolas
  Papernot. 2021.
\newblock \href
  {https://www.usenix.org/conference/usenixsecurity21/presentation/jia}
  {Entangled watermarks as a defense against model extraction}.
\newblock In \emph{30th {USENIX} Security Symposium, {USENIX} Security 2021,
  August 11-13, 2021}, pages 1937--1954. {USENIX} Association.

\bibitem[{Kirchenbauer et~al.(2023)Kirchenbauer, Geiping, Wen, Katz, Miers, and
  Goldstein}]{wm6}
John Kirchenbauer, Jonas Geiping, Yuxin Wen, Jonathan Katz, Ian Miers, and Tom
  Goldstein. 2023.
\newblock \href {https://proceedings.mlr.press/v202/kirchenbauer23a.html} {A
  watermark for large language models}.
\newblock In \emph{Proceedings of the 40th International Conference on Machine
  Learning}, volume 202 of \emph{Proceedings of Machine Learning Research},
  pages 17061--17084. PMLR.

\bibitem[{Kirk et~al.(2023)Kirk, Mediratta, Nalmpantis, Luketina, Hambro,
  Grefenstette, and Raileanu}]{tldr}
Robert Kirk, Ishita Mediratta, Christoforos Nalmpantis, Jelena Luketina, Eric
  Hambro, Edward Grefenstette, and Roberta Raileanu. 2023.
\newblock Understanding the effects of rlhf on llm generalisation and
  diversity.
\newblock \emph{arXiv preprint arXiv:2310.06452}.

\bibitem[{Korbak et~al.(2022)Korbak, Elsahar, Kruszewski, and
  Dymetman}]{rl-deduct4}
Tomasz Korbak, Hady Elsahar, Germ{\'{a}}n Kruszewski, and Marc Dymetman. 2022.
\newblock \href
  {http://papers.nips.cc/paper\_files/paper/2022/hash/67496dfa96afddab795530cc7c69b57a-Abstract-Conference.html}
  {On reinforcement learning and distribution matching for fine-tuning language
  models with no catastrophic forgetting}.
\newblock In \emph{Advances in Neural Information Processing Systems 35: Annual
  Conference on Neural Information Processing Systems 2022, NeurIPS 2022, New
  Orleans, LA, USA, November 28 - December 9, 2022}.

\bibitem[{Krishna et~al.(2020)Krishna, Tomar, Parikh, Papernot, and
  Iyyer}]{mea-bertapi}
Kalpesh Krishna, Gaurav~Singh Tomar, Ankur~P. Parikh, Nicolas Papernot, and
  Mohit Iyyer. 2020.
\newblock \href {https://openreview.net/forum?id=Byl5NREFDr} {Thieves on sesame
  street! model extraction of bert-based apis}.
\newblock In \emph{8th International Conference on Learning Representations,
  {ICLR} 2020, Addis Ababa, Ethiopia, April 26-30, 2020}. OpenReview.net.

\bibitem[{Lee et~al.(2023)Lee, Phatale, Mansoor, Mesnard, Ferret, Lu, Bishop,
  Hall, Carbune, Rastogi, and Prakash}]{rlaif}
Harrison Lee, Samrat Phatale, Hassan Mansoor, Thomas Mesnard, Johan Ferret,
  Kellie Lu, Colton Bishop, Ethan Hall, Victor Carbune, Abhinav Rastogi, and
  Sushant Prakash. 2023.
\newblock \href {https://arxiv.org/abs/2309.00267} {Rlaif: Scaling
  reinforcement learning from human feedback with ai feedback}.
\newblock \emph{Preprint}, arXiv:2309.00267.

\bibitem[{Li et~al.(2023{\natexlab{a}})Li, Cheng, Li, Du, Zhao, and
  Liu}]{model-wm3}
Peixuan Li, Pengzhou Cheng, Fangqi Li, Wei Du, Haodong Zhao, and Gongshen Liu.
  2023{\natexlab{a}}.
\newblock \href {https://doi.org/10.1609/aaai.v37i12.26750} {Plmmark: A secure
  and robust black-box watermarking framework for pre-trained language models}.
\newblock \emph{Proceedings of the AAAI Conference on Artificial Intelligence},
  37(12):14991--14999.

\bibitem[{Li et~al.(2023{\natexlab{b}})Li, Wang, Ma, Liu, Wang, Wu, Gao, and
  Liu}]{mea-code}
Zongjie Li, Chaozheng Wang, Pingchuan Ma, Chaowei Liu, Shuai Wang, Daoyuan Wu,
  Cuiyun Gao, and Yang Liu. 2023{\natexlab{b}}.
\newblock \href {https://arxiv.org/abs/2303.03012} {On extracting specialized
  code abilities from large language models: A feasibility study}.
\newblock \emph{Preprint}, arXiv:2303.03012.

\bibitem[{Liang et~al.(2025{\natexlab{a}})Liang, Hu, Ye, Xiao, and Li}]{prompt}
Zi~Liang, Haibo Hu, Qingqing Ye, Yaxin Xiao, and Haoyang Li.
  2025{\natexlab{a}}.
\newblock \href {https://arxiv.org/abs/2408.02416} {Why are my prompts leaked?
  unraveling prompt extraction threats in customized large language models}.
\newblock \emph{Preprint}, arXiv:2408.02416.

\bibitem[{Liang et~al.(2025{\natexlab{b}})Liang, Wang, Zhang, Hu, Zhang, Ye,
  Xu, Xiao, Zhang, and Cui}]{isaac}
Zi~Liang, Pinghui Wang, Ruofei Zhang, Haibo Hu, Shuo Zhang, Qingqing Ye, Nuo
  Xu, Yaxin Xiao, Chen Zhang, and Lizhen Cui. 2025{\natexlab{b}}.
\newblock \href {https://doi.org/10.1609/aaai.v39i26.34957} {Exploring
  intrinsic alignments within text corpus}.
\newblock \emph{Proceedings of the AAAI Conference on Artificial Intelligence},
  39(26):27455--27463.

\bibitem[{Lin et~al.(2020)Lin, Zhou, Shen, Zhou, Bhagavatula, Choi, and
  Ren}]{commongen}
Bill~Yuchen Lin, Wangchunshu Zhou, Ming Shen, Pei Zhou, Chandra Bhagavatula,
  Yejin Choi, and Xiang Ren. 2020.
\newblock \href {https://doi.org/10.18653/v1/2020.findings-emnlp.165}
  {{C}ommon{G}en: A constrained text generation challenge for generative
  commonsense reasoning}.
\newblock In \emph{Findings of the Association for Computational Linguistics:
  EMNLP 2020}, pages 1823--1840, Online. Association for Computational
  Linguistics.

\bibitem[{Lin(2004)}]{rouge}
Chin-Yew Lin. 2004.
\newblock \href {https://aclanthology.org/W04-1013} {{ROUGE}: A package for
  automatic evaluation of summaries}.
\newblock In \emph{Text Summarization Branches Out}, pages 74--81, Barcelona,
  Spain. Association for Computational Linguistics.

\bibitem[{Lin et~al.(2021)Lin, Hilton, and Evans}]{tqa}
Stephanie Lin, Jacob Hilton, and Owain Evans. 2021.
\newblock \href {https://arxiv.org/abs/2109.07958} {Truthfulqa: Measuring how
  models mimic human falsehoods}.
\newblock \emph{Preprint}, arXiv:2109.07958.

\bibitem[{Lv et~al.(2024)Lv, Ma, Chen, Zhou, Zhang, Liang, Zhu, Li, and
  Zhang}]{mea-defender}
P.~Lv, H.~Ma, K.~Chen, J.~Zhou, S.~Zhang, R.~Liang, S.~Zhu, P.~Li, and
  Y.~Zhang. 2024.
\newblock \href {https://doi.org/10.1109/SP54263.2024.00099} {Mea-defender: A
  robust watermark against model extraction attack}.
\newblock In \emph{2024 IEEE Symposium on Security and Privacy (SP)}, pages
  102--102, Los Alamitos, CA, USA. IEEE Computer Society.

\bibitem[{Meng et~al.(2024)Meng, Xia, and Chen}]{simpo}
Yu~Meng, Mengzhou Xia, and Danqi Chen. 2024.
\newblock Simpo: Simple preference optimization with a reference-free reward.
\newblock \emph{arXiv preprint arXiv:2405.14734}.

\bibitem[{Myung(2003)}]{mle}
In~Myung. 2003.
\newblock \href {https://doi.org/10.1016/S0022-2496(02)00028-7} {Tutorial on
  maximum likelihood estimation}.
\newblock \emph{Journal of Mathematical Psychology}, 47:90--100.

\bibitem[{OpenAI(2024)}]{openai-doc}
OpenAI. 2024.
\newblock Openai api reference documentation: chat.
\newblock [Online].
\newblock \url{https://platform.openai.com/docs/api-reference/chat}.

\bibitem[{Ouyang et~al.(2022)Ouyang, Wu, Jiang, Almeida, Wainwright, Mishkin,
  Zhang, Agarwal, Slama, and et~al.}]{instructGPT}
Long Ouyang, Jeff Wu, Xu~Jiang, Diogo Almeida, Carroll~L. Wainwright, Pamela
  Mishkin, Chong Zhang, Sandhini Agarwal, Katarina Slama, and Alex~Ray et~al.
  2022.
\newblock Training language models to follow instructions with human feedback.
\newblock \emph{ArXiv}, abs/2203.02155.

\bibitem[{Papernot et~al.(2017)Papernot, McDaniel, Goodfellow, Jha, Celik, and
  Swami}]{mea-tra-prac}
Nicolas Papernot, Patrick McDaniel, Ian Goodfellow, Somesh Jha, Z~Berkay Celik,
  and Ananthram Swami. 2017.
\newblock Practical black-box attacks against machine learning.
\newblock In \emph{Proceedings of the 2017 ACM on Asia conference on computer
  and communications security}, pages 506--519.

\bibitem[{Papineni et~al.(2002)Papineni, Roukos, Ward, and Zhu}]{bleu}
Kishore Papineni, Salim Roukos, Todd Ward, and Wei-Jing Zhu. 2002.
\newblock \href {https://doi.org/10.3115/1073083.1073135} {{B}leu: a method for
  automatic evaluation of machine translation}.
\newblock In \emph{Proceedings of the 40th Annual Meeting of the Association
  for Computational Linguistics}, pages 311--318, Philadelphia, Pennsylvania,
  USA. Association for Computational Linguistics.

\bibitem[{Peng et~al.(2023)Peng, Yi, Wu, Wu, Bin~Zhu, Lyu, Jiao, Xu, Sun, and
  Xie}]{model-wm1}
Wenjun Peng, Jingwei Yi, Fangzhao Wu, Shangxi Wu, Bin Bin~Zhu, Lingjuan Lyu,
  Binxing Jiao, Tong Xu, Guangzhong Sun, and Xing Xie. 2023.
\newblock \href {https://doi.org/10.18653/v1/2023.acl-long.423} {Are you
  copying my model? protecting the copyright of large language models for
  {E}aa{S} via backdoor watermark}.
\newblock In \emph{Proceedings of the 61st Annual Meeting of the Association
  for Computational Linguistics (Volume 1: Long Papers)}, pages 7653--7668,
  Toronto, Canada. Association for Computational Linguistics.

\bibitem[{Peng et~al.(2019)Peng, Kumar, Zhang, and Levine}]{rl-deduct2}
Xue~Bin Peng, Aviral Kumar, Grace Zhang, and Sergey Levine. 2019.
\newblock \href {https://arxiv.org/abs/1910.00177} {Advantage-weighted
  regression: Simple and scalable off-policy reinforcement learning}.
\newblock \emph{CoRR}, abs/1910.00177.

\bibitem[{Perez et~al.(2023)Perez, Ringer, Lukosiute, Nguyen, Chen, Heiner,
  Pettit, Olsson, Kundu, and et~al.}]{rlhf3}
Ethan Perez, Sam Ringer, Kamile Lukosiute, Karina Nguyen, Edwin Chen, Scott
  Heiner, Craig Pettit, Catherine Olsson, Sandipan Kundu, and et~al. 2023.
\newblock \href {https://doi.org/10.18653/V1/2023.FINDINGS-ACL.847}
  {Discovering language model behaviors with model-written evaluations}.
\newblock In \emph{{ACL} 2023,}, pages 13387--13434. Association for
  Computational Linguistics.

\bibitem[{Peters and Schaal(2007)}]{rl-deduct1}
Jan Peters and Stefan Schaal. 2007.
\newblock \href {https://doi.org/10.1145/1273496.1273590} {Reinforcement
  learning by reward-weighted regression for operational space control}.
\newblock In \emph{Machine Learning, Proceedings of the Twenty-Fourth
  International Conference {(ICML} 2007), Corvallis, Oregon, USA, June 20-24,
  2007}, volume 227 of \emph{{ACM} International Conference Proceeding Series},
  pages 745--750. {ACM}.

\bibitem[{PyTorch(2024)}]{torch}
PyTorch. 2024.
\newblock Softmax doesn’t work directly with nllloss, which expects the log
  to be computed between the softmax and itself. use log\_softmax instead
  (it’s faster and has better numerical properties).
\newblock [Online].
\newblock
  \url{https://pytorch.org/docs/stable/generated/torch.nn.functional.softmax.html#torch.nn.functional.softmax}.

\bibitem[{Rafailov et~al.(2023)Rafailov, Sharma, Mitchell, Manning, Ermon, and
  Finn}]{dpo}
Rafael Rafailov, Archit Sharma, Eric Mitchell, Christopher~D. Manning, Stefano
  Ermon, and Chelsea Finn. 2023.
\newblock \href
  {http://papers.nips.cc/paper\_files/paper/2023/hash/a85b405ed65c6477a4fe8302b5e06ce7-Abstract-Conference.html}
  {Direct preference optimization: Your language model is secretly a reward
  model}.
\newblock In \emph{Advances in Neural Information Processing Systems 36: Annual
  Conference on Neural Information Processing Systems 2023, NeurIPS 2023, New
  Orleans, LA, USA, December 10 - 16, 2023}.

\bibitem[{Rafi et~al.(2022)Rafi, Feng, and Jeon}]{mea-vendor}
Mujahid~Al Rafi, Yuan Feng, and Hyeran Jeon. 2022.
\newblock \href {https://doi.org/10.48550/ARXIV.2207.09539} {Revealing secrets
  from pre-trained models}.
\newblock \emph{CoRR}, abs/2207.09539.

\bibitem[{Saad and Solla(1995)}]{mea-the-dyn}
David Saad and Sara Solla. 1995.
\newblock Dynamics of on-line gradient descent learning for multilayer neural
  networks.
\newblock \emph{Advances in neural information processing systems}, 8.

\bibitem[{Schulman et~al.(2015)Schulman, Levine, Moritz, Jordan, and
  Abbeel}]{trpo}
John Schulman, Sergey Levine, Philipp Moritz, Michael~I. Jordan, and Pieter
  Abbeel. 2015.
\newblock \href {https://arxiv.org/abs/1502.05477} {Trust region policy
  optimization}.
\newblock \emph{CoRR}, abs/1502.05477.

\bibitem[{Schulman et~al.(2017)Schulman, Wolski, Dhariwal, Radford, and
  Klimov}]{ppo}
John Schulman, Filip Wolski, Prafulla Dhariwal, Alec Radford, and Oleg Klimov.
  2017.
\newblock \href {https://arxiv.org/abs/1707.06347} {Proximal policy
  optimization algorithms}.
\newblock \emph{Preprint}, arXiv:1707.06347.

\bibitem[{Shazeer et~al.(2017)Shazeer, Mirhoseini, Maziarz, Davis, Le, Hinton,
  and Dean}]{moe}
Noam Shazeer, Azalia Mirhoseini, Krzysztof Maziarz, Andy Davis, Quoc Le,
  Geoffrey Hinton, and Jeff Dean. 2017.
\newblock Outrageously large neural networks: The sparsely-gated
  mixture-of-experts layer.
\newblock \emph{arXiv preprint arXiv:1701.06538}.

\bibitem[{Sun et~al.(2022)Sun, Xu, Deng, Cheng, Zheng, Zhou, Peng, Zhu, and
  Huang}]{diasafety}
Hao Sun, Guangxuan Xu, Jiawen Deng, Jiale Cheng, Chujie Zheng, Hao Zhou, Nanyun
  Peng, Xiaoyan Zhu, and Minlie Huang. 2022.
\newblock On the safety of conversational models: Taxonomy, dataset, and
  benchmark.
\newblock In \emph{Findings of ACL 2022}.

\bibitem[{Taori et~al.(2023)Taori, Gulrajani, Zhang, Dubois, Li, Guestrin,
  Liang, and Hashimoto}]{alpaca}
Rohan Taori, Ishaan Gulrajani, Tianyi Zhang, Yann Dubois, Xuechen Li, Carlos
  Guestrin, Percy Liang, and Tatsunori~B. Hashimoto. 2023.
\newblock Stanford alpaca: An instruction-following llama model.
\newblock \url{https://github.com/tatsu-lab/stanford_alpaca}.

\bibitem[{Tian(2020)}]{mea-the-stu}
Yuandong Tian. 2020.
\newblock Student specialization in deep rectified networks with finite width
  and input dimension.
\newblock In \emph{International Conference on Machine Learning}, pages
  9470--9480. PMLR.

\bibitem[{Tram{\`e}r et~al.(2016)Tram{\`e}r, Zhang, Juels, Reiter, and
  Ristenpart}]{mea-tra-steal}
Florian Tram{\`e}r, Fan Zhang, Ari Juels, Michael~K Reiter, and Thomas
  Ristenpart. 2016.
\newblock Stealing machine learning models via prediction apis.
\newblock In \emph{25th USENIX security symposium (USENIX Security 16)}, pages
  601--618.

\bibitem[{Wallace et~al.(2020)Wallace, Stern, and Song}]{mea-wmt}
Eric Wallace, Mitchell Stern, and Dawn Song. 2020.
\newblock \href {https://doi.org/10.18653/V1/2020.EMNLP-MAIN.446} {Imitation
  attacks and defenses for black-box machine translation systems}.
\newblock In \emph{Proceedings of the 2020 Conference on Empirical Methods in
  Natural Language Processing, {EMNLP} 2020, Online, November 16-20, 2020},
  pages 5531--5546. Association for Computational Linguistics.

\bibitem[{Xiao et~al.(2022)Xiao, Ye, Hu, Zheng, Fang, and Shi}]{mea-12}
Yaxin Xiao, Qingqing Ye, Haibo Hu, Huadi Zheng, Chengfang Fang, and Jie Shi.
  2022.
\newblock \href
  {https://proceedings.neurips.cc/paper_files/paper/2022/file/4241c27d3161c7a7064bfc1a6e539563-Paper-Conference.pdf}
  {Mexmi: Pool-based active model extraction crossover membership inference}.
\newblock In \emph{Advances in Neural Information Processing Systems},
  volume~35, pages 10203--10216. Curran Associates, Inc.

\bibitem[{Xu et~al.(2022)Xu, He, Lyu, Qu, and Haffari}]{mea-ensemble}
Qiongkai Xu, Xuanli He, Lingjuan Lyu, Lizhen Qu, and Gholamreza Haffari. 2022.
\newblock \href {https://aclanthology.org/2022.coling-1.251} {Student surpasses
  teacher: Imitation attack for black-box {NLP} apis}.
\newblock In \emph{Proceedings of the 29th International Conference on
  Computational Linguistics, {COLING} 2022, Gyeongju, Republic of Korea,
  October 12-17, 2022}, pages 2849--2860. International Committee on
  Computational Linguistics.

\bibitem[{Yu et~al.(2018)Yu, Zhang, Yang, Yasunaga, Wang, Li, Ma, Li, Yao,
  Roman et~al.}]{spider}
Tao Yu, Rui Zhang, Kai Yang, Michihiro Yasunaga, Dongxu Wang, Zifan Li, James
  Ma, Irene Li, Qingning Yao, Shanelle Roman, et~al. 2018.
\newblock Spider: A large-scale human-labeled dataset for complex and
  cross-domain semantic parsing and text-to-sql task.
\newblock \emph{arXiv preprint arXiv:1809.08887}.

\bibitem[{Zhang et~al.(2022)Zhang, Roller, Goyal, Artetxe, Chen, Chen, Dewan,
  Diab, Li, Lin, Mihaylov, Ott, Shleifer, Shuster, Simig, Koura, Sridhar, Wang,
  and Zettlemoyer}]{opt}
Susan Zhang, Stephen Roller, Naman Goyal, Mikel Artetxe, Moya Chen, Shuohui
  Chen, Christopher Dewan, Mona Diab, Xian Li, Xi~Victoria Lin, Todor Mihaylov,
  Myle Ott, Sam Shleifer, Kurt Shuster, Daniel Simig, Punit~Singh Koura, Anjali
  Sridhar, Tianlu Wang, and Luke Zettlemoyer. 2022.
\newblock \href {https://arxiv.org/abs/2205.01068} {Opt: Open pre-trained
  transformer language models}.
\newblock \emph{Preprint}, arXiv:2205.01068.

\bibitem[{Zhang et~al.(2020)Zhang, Kishore, Wu, Weinberger, and
  Artzi}]{bertscore}
Tianyi Zhang, Varsha Kishore, Felix Wu, Kilian~Q. Weinberger, and Yoav Artzi.
  2020.
\newblock \href {https://openreview.net/forum?id=SkeHuCVFDr} {Bertscore:
  Evaluating text generation with {BERT}}.
\newblock In \emph{8th International Conference on Learning Representations,
  {ICLR} 2020, Addis Ababa, Ethiopia, April 26-30, 2020}. OpenReview.net.

\bibitem[{Zhang et~al.(2025)Zhang, Hu, Ye, Bai, and Zheng}]{mea-13}
Xinwei Zhang, Haibo Hu, Qingqing Ye, Li~Bai, and Huadi Zheng. 2025.
\newblock \href {https://doi.org/10.1145/3696410.3714894} {Mer-inspector:
  Assessing model extraction risks from an attack-agnostic perspective}.
\newblock In \emph{Proceedings of the ACM on Web Conference 2025}, WWW '25,
  page 4300–4315, New York, NY, USA. Association for Computing Machinery.

\bibitem[{Zhao et~al.(2022)Zhao, Li, and Wang}]{wm3}
Xuandong Zhao, Lei Li, and Yu{-}Xiang Wang. 2022.
\newblock \href {https://doi.org/10.18653/V1/2022.FINDINGS-EMNLP.370}
  {Distillation-resistant watermarking for model protection in {NLP}}.
\newblock In \emph{Findings of the Association for Computational Linguistics:
  {EMNLP} 2022, Abu Dhabi, United Arab Emirates, December 7-11, 2022}, pages
  5044--5055. Association for Computational Linguistics.

\bibitem[{Zhao et~al.(2023)Zhao, Wang, and Li}]{wm7}
Xuandong Zhao, Yu-Xiang Wang, and Lei Li. 2023.
\newblock Protecting language generation models via invisible watermarking.
\newblock In \emph{Proceedings of the 40th International Conference on Machine
  Learning}, ICML'23. JMLR.org.

\bibitem[{Zheng et~al.(2019)Zheng, Ye, Hu, Fang, and Shi}]{mea-11}
Huadi Zheng, Qingqing Ye, Haibo Hu, Chengfang Fang, and Jie Shi. 2019.
\newblock Bdpl: A boundary differentially private layer against machine
  learning model extraction attacks.
\newblock In \emph{Computer Security -- ESORICS 2019}, pages 66--83, Cham.
  Springer International Publishing.

\bibitem[{Zhong et~al.(2017)Zhong, Xiong, and Socher}]{wikisql}
Victor Zhong, Caiming Xiong, and Richard Socher. 2017.
\newblock Seq2sql: Generating structured queries from natural language using
  reinforcement learning.
\newblock \emph{CoRR}, abs/1709.00103.

\bibitem[{Zhou et~al.(2021)Zhou, Ge, and Jin}]{mea-the-a}
Mo~Zhou, Rong Ge, and Chi Jin. 2021.
\newblock A local convergence theory for mildly over-parameterized two-layer
  neural network.
\newblock In \emph{Conference on Learning Theory}, pages 4577--4632. PMLR.

\end{thebibliography}
